\documentclass[pre,twocolumn,aps,showpacs,amsmath,amssymb,superscriptaddress]{revtex4-1}

\usepackage{amsmath}
\usepackage{amssymb}
\usepackage{amsbsy}
\usepackage{graphicx}
\usepackage{color}
\usepackage{mathtools}

\def \equi#1{\mathrel{\mathop{\kern 0pt\sim}\limits_{#1}}} 

\usepackage{enumerate}

\newcommand{\moy}[1]{\left\langle #1 \right\rangle}

\newcommand{\ex}[1]{\mathrm{e}^{#1}}
\newcommand{\epu}[0]{\ee_1}
\newcommand{\emu}[0]{\ee_{-1}}

\newcommand{\dd}[0]{\mathrm{d}}
\newcommand{\ii}[0]{\mathrm{i}}

\newcommand{\zz}[0]{\mathbf{0}}
\newcommand{\ZZ}[0]{\boldsymbol{Z}}
\newcommand{\FF}[2]{\widehat{F} \left( #1|#2\,;\xi \right)}
\newcommand{\ee}[0]{\boldsymbol{e}}
\newcommand{\rr}[0]{\boldsymbol{r}}
\newcommand{\ep}[0]{\epsilon}
\newcommand{\Ft}[2]{{F}_t ( #1|#2 )}

\begin{document}

\title{Diffusion and subdiffusion of interacting particles on comb-like structures}

\author{O. B\'enichou}
\affiliation{Sorbonne Universit\'es, UPMC Univ Paris 06, UMR 7600, LPTMC, F-75005, Paris, France}
\affiliation{CNRS, UMR 7600, Laboratoire de Physique Th\'{e}orique de la Mati\`{e}re Condens\'{e}e, F-75005, Paris, France}

\author{P. Illien}
\affiliation{Sorbonne Universit\'es, UPMC Univ Paris 06, UMR 7600, LPTMC, F-75005, Paris, France}
\affiliation{CNRS, UMR 7600, Laboratoire de Physique Th\'{e}orique de la Mati\`{e}re Condens\'{e}e, F-75005, Paris, France}

\author{G. Oshanin}
\affiliation{Sorbonne Universit\'es, UPMC Univ Paris 06, UMR 7600, LPTMC, F-75005, Paris, France}
\affiliation{CNRS, UMR 7600, Laboratoire de Physique Th\'{e}orique de la Mati\`{e}re Condens\'{e}e, F-75005, Paris, France}

\author{A. Sarracino}
\affiliation{Sorbonne Universit\'es, UPMC Univ Paris 06, UMR 7600, LPTMC, F-75005, Paris, France}
\affiliation{CNRS, UMR 7600, Laboratoire de Physique Th\'{e}orique de la Mati\`{e}re Condens\'{e}e, F-75005, Paris, France}

\author{R. Voituriez}
\affiliation{Sorbonne Universit\'es, UPMC Univ Paris 06, UMR 7600, LPTMC, F-75005, Paris, France}
\affiliation{CNRS, UMR 7600, Laboratoire de Physique Th\'{e}orique de la Mati\`{e}re Condens\'{e}e, F-75005, Paris, France}

\date{\today}

\begin{abstract}
  We study the dynamics of a tracer particle (TP) on a comb lattice
  populated by randomly moving hard-core particles in the dense limit. We first
  consider the case where the TP is constrained to move on the backbone
  of the comb only, and, in the limit of high density of particles, we
  present exact analytical results for the cumulants of the TP
  position, showing a subdiffusive behavior $\sim t^{3/4}$.  At longer
  times, a second regime is observed, where standard diffusion is
  recovered, with a surprising non analytical dependence of the
  diffusion coefficient on the particle density.  When the TP is
  allowed to visit the teeth of the comb, based on a mean-field-like
  Continuous Time Random Walk description, we unveil a rich and
  complex scenario, with several successive subdiffusive regimes,
  resulting from the coupling between the inhomogeneous comb geometry
  and particle interactions. Remarkably, the presence of hard-core
  interactions speeds up the TP motion along the backbone of the
  structure in all regimes.
\end{abstract}

%\pacs{83.10.-y,05.40.Fb,83.10.Pp}

\maketitle

%\emph{Introduction.}--

\begin{figure}
\begin{center}
\includegraphics[width=0.8\columnwidth,clip=true]{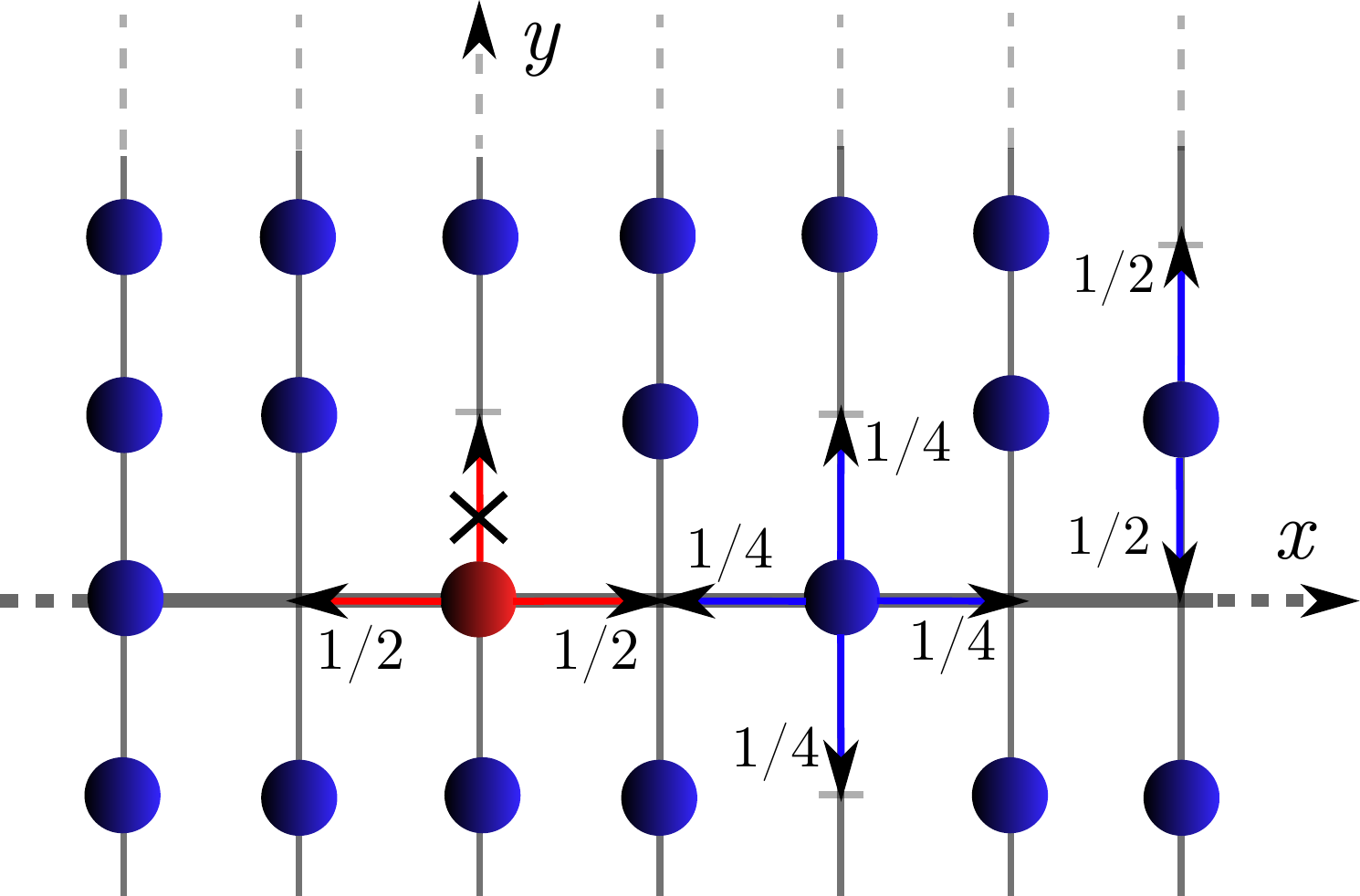}
\caption{Geometry of the system. The $x$-axis is designated as the
  \textit{backbone} of the lattice, whereas the orthogonal lines are
  designated as the \textit{teeth}. Jump rules of the particles in the
  case when the tracer particle (in red) is constrained to move on the
  backbone are given.}
\label{model}
\end{center}
\end{figure}

Subdiffusive motion of tracer particles in crowded media, e.g.
biological cells, is widespread. Among the possible
microscopic scenarios leading to this sublinear growth with time of
the mean square displacement (MSD), the existence of geometric
constraints related to the complexity of the environment plays 
an important role~\cite{Metzler2000,Condamin2008}.  In
this context, the comb model (see Fig.~\ref{model}), in which a single
particle moves on a two-dimensional space with the
constraint that steps in the $x$ direction are only allowed when the
$y$ coordinate of the particle positions is zero, has attracted
considerable attention because of its simplicity and ability to
reproduce subdiffusive behaviors of disordered
systems~\cite{Ben-avraham}.

Comb-like structures have indeed been introduced as a
first step to model diffusion in more complicated fractal structures
like percolation clusters, the backbone and teeth of the comb
representing the quasilinear structure and dangling ends of
percolation clusters~\cite{Weiss1986}. The particle can spend a long
time exploring a tooth, which results in a subdiffusive motion along
the backbone with $\langle x^2(t)\rangle \propto t^\alpha$
with $\alpha=1/2$. Since, numerous results have been obtained for this
model~\cite{Arkhincheev1991,Burioni2005a,Frauenrath2005a,Iomin2011,Villamaina2011,Mendez2013,Lenzi2013,Agliari2014},
including the determination of the occupation time
statistics~\cite{Rebenshtok2013}, of mean first-passage times between
two nodes of a finite comb~\cite{Agliari2015} or the case of
fractional Brownian walks on comb-like structures~\cite{Ribeiro2014}.

In parallel, the comb model has been invoked to account for transport
in real systems like spiny dendrites~\cite{Mendez2013}, diffusion of
cold atoms~\cite{Sagi2012a} and mainly diffusion in crowded media like
cells~\cite{Hofling2013}. However, all existing studies have focused
on single-particle diffusion, and interactions between particles have
up to now been completely left aside. As an elementary model for
diffusion of particles under short-range repulsive forces, we
consider here excluded-volume interactions (EVIs) and focus on their
impact on tracer dynamics on comb-like structures.

From a theoretical point of view, lattice systems of interacting
particles represent a prototypical model in statistical physics that
has generated a huge number of works both in the
physical~\cite{Chou2011,Levine2001} and mathematical
literature~\cite{Spohn1991}. The effect of EVIs in {\it homogeneous}
lattices is well known~\cite{Ben-avraham}. In dimension $d\ge2$,
tracer diffusion has been shown to remain normal, with a non trivial
diffusion coefficient resulting from many-body interactions and well
approximated by the Nakazato-Kitahara approach~\cite{Nakazato1980}. In
a ``single-file" geometry, where particles cannot bypass each other,
the impact of EVIs is stronger and results in a subdiffusive behavior
$\langle x^2(t)\rangle \propto t^\beta$ with
$\beta=1/2$~\cite{Harris1965,Levitt1973,Arratia1983,Leibovich2013,Krapivsky2014,Hegde2014}. In
this context, determining the effect of EVIs on systems with {\it
  geometrical constraints} appears an important question which does
not seem to have received much attention. Notable exceptions
are~\cite{Burioni2002}, where two particles only are involved,~\cite{Sanders2014}, which involves single-file motion with
waiting times and thus does not consider explicitly geometric
constraints, and~\cite{Amitrano}, where tracer diffusion on DLA
clusters was studied numerically and found to be not modified by the
presence of EVIs.  In this Letter, we show that, in contrast, EVIs
deeply modify tracer diffusion on comb-like structures. Focusing on
the high density limit, we show analytically that, due to a subtle
interplay between the inhomogeneous comb geometry and crowding
effects, the dynamics displays several regimes of anomalous
diffusion. We find in particular that, surprisingly, the presence of
EVIs can speed up tracer diffusion along the backbone of the
structure.
 
\emph{Model.}-- We consider the $2$-dimensional comb $\mathbf{C}_2$,
which is a subgraph of $\mathbb{Z}^2$ obtained by removing all the
edges parallel to the $x$-axis, except from the $x$-axis itself. This
lattice is populated by $N$ hard-core particles with average density
$\rho=N/V$, where $V$ is the number of sites. Each particle performs a
symmetric nearest-neighbor random walk, constrained by hard-core
interactions. We add a tracer particle (TP) at the origin,
performing a symmetric nearest-neighbor random walk, and focus on its
dynamics in the dense limit, where the vacancy density
$\rho_0=1-\rho\ll 1$.  In this limit, it is more convenient to describe the vacancy dynamics
instead of describing the dynamics of all the particles.  We assume
here that, at each time step, each vacancy exchanges its position with
one of the neighboring particles, with jump probabilities that depend
on the position on the lattice; see the Supplemental
Material (SM)~\cite{SM} for the explicit definition of those evolution
rules.

\emph{Case of a TP restricted to the backbone}-- We first assume that
the TP (and only the TP) is constrained to move on the backbone. This particular case is important for several reasons. (i)
It mimics the case where the tracer is different from the bath
particles, and is not able to visit the teeth of the comb. (ii) At the
theoretical level, it appears as an extension of the famous
single-file geometry in which, due do the possibility for the bath
particles to visit the teeth, the particles can bypass each other. An
interesting question is to know if the dynamics is still anomalous in
this case, and if so, with which exponent; (iii) finally, as shown
below, solving this auxiliary problem will allow us to determine the
dynamics of the TP in the general case where the TP can access the
teeth of the comb.

Let $X_t$ be the random variable characterizing the position of the TP
along the backbone at time $t$. We aim at computing the cumulants of
arbitrary order $n$ of this variable, denoted by
$\kappa^{(n)}(t)$. These quantities are generated by the cumulant
generating function (CGF) $\Psi_t(k)\equiv\ln\moy{\ex{ikX_t}}=\ln
\left[ \widetilde{P}_t(k)\right]$, where
$\widetilde{P}_t(k)=\sum_{X}e^{ikX} P_t(X)$ is the Fourier transform
of the probability $P_t(X)$ to find the TP at position ${X}$ at time
$t$.  Following the method developed in~\cite{Brummelhuis1989a} and
recently used to study driven diffusion in one-dimensional
geometries~\cite{Illien2013a}, we first consider the case where there
is a single vacancy on the lattice.  Let $P_t^{(1)}(X|\ZZ)$ be the
probability to find the TP at position $X$ at time $t$ knowing that
the vacancy started from site $\ZZ$. Summing over all the passages of
the vacancy to the TP location, one gets:
\begin{eqnarray}
  &&P_t^{(1)}(X|\ZZ)=\delta_{ X,{ 0}}\left(1-\sum_{j=0}^t F_j(\zz|\ZZ)\right) \nonumber \\
  &+&\sum_{p=1}^{+\infty}\sum_{m_1,\ldots,m_p=1}^{+\infty} \sum_{m_{p+1}=0}^{+\infty}\delta_{m_1+\ldots+m_{p+1},t}\delta_{X,\frac{{\rm sgn}(\ZZ\cdot\mathbf{e_1})+(-1)^{p+1}}{2}}   \nonumber\\
  &\times& \left(1-\sum_{j=0}^{m_{p+1}}F_j(\zz|(-1)^p\mathbf{e_1})\right)\nonumber \\ 
  &\times& F_{m_p}(\zz|(-1)^{p+1}\mathbf{e_1}) \ldots F_{m_1}(\zz|\ZZ),
\label{Ptr}
\end{eqnarray}
where $F_t(\zz|\ZZ)$ is the probability for the vacancy to reach the
origin for the first time at time $t$, knowing that it started from
site $\ZZ$, and $\mathbf{e_1}$ stands for the unit vector in the $x$
direction. The first term in the right-hand side of Eq.~(\ref{Ptr})
represents the event that at time $t$, the TP has not been visited by
any vacancy, while the second one results from a partition both on the
number $p$ of visits and waiting times $m_i$ between visits of the TP
by the vacancy.  Computing the generating function associated with
this propagator $ \widehat{p}_{\pm 1}
(X;\xi)\equiv\widehat{P}^{(1)}(X|\pm \epu ; \xi)$, where
$\widehat{\phi}(\xi)$ denotes the discrete Laplace transform
$\widehat{\phi}(\xi)\equiv \sum_{t=0}^\infty\phi_t\xi^t$, and noticing
that for symmetry reasons
$\widehat{F}(\zz|\epu;\xi)=\widehat{F}(\zz|-\epu;\xi)\equiv\widehat{F}_{1}$,
one gets
 \begin{equation}
   \label{Psingle}
   \widehat{p}_{\pm 1} (X;\xi) = \frac{\delta_{X,0}(1-\widehat{F}_{  1})+\delta_{X,\pm 1}\widehat{F}_{ 1}(1-\widehat{F}_{ 1})}{(1-{\widehat{F}_{ 1}}^2)(1-\xi)}.
\end{equation}

We then study the case where the concentration of vacancies on the
lattice $\rho_0$ is finite but very small. For clarity, we first
assume that the lattice has a finite number of sites $N$, and that it
is populated by $M$ vacancies, so that 
$M=\rho_0 N$. Consequently, the probability $P_t(X|\{\ZZ_j\})$ to find
the TP at position $X$ as a result of its interactions with $M$
vacancies initially located at sites $\ZZ_1,\cdots,\ZZ_M$, is given by
\begin{equation}
P_t(X|\{\ZZ_j\})=\sum_{\ZZ_1,\cdots,\ZZ_M} \delta_{X,Y_1+\cdots+Y_M} P_t(\{ Y_j \}|\{\ZZ_j \}),
\end{equation}
where $P_t(\{ Y_j \}|\{\ZZ_j \})$ is the conditional probability that
during the time interval $t$ the TP moved of a distance $Y_j$ due to
its interactions with the $j$-th vacancy. To leading order in
$\rho_0$, the vacancies contribute independently to the displacement
of the TP, so that in Fourier variable,
$\widetilde{P}^{(M)}_t(k)=\left[ \widetilde{P}_t^{(1)}(k) \right]^M$,
where $\widetilde{P}^{(j)}_t(k)$ is the Fourier transform of the
probability distribution to find the TP at position $X$ at time $t$,
knowing that there are $j$ vacancies on the lattice, and averaged over
the initial position of the vacancies, which is assumed to be
uniform. As shown below, the choice of the initial distribution of the
vacancies may actually have a dramatic effect on the behavior of the
TP.

Finally, in the thermodynamic limit in which $N,M\to\infty$ with fixed
$\rho_0=M/N$, and using Eqs.~(\ref{Psingle}) we get the Fourier
Laplace transform of the CGF
\begin{equation}
\label{Psidef}
{\widehat{\Psi}(k;\xi)} \underset{\rho_0 \to 0}{\sim} - 2 \rho_0 \frac{ H(\xi)}{(1-\xi)(1+\widehat{F}_1)} (1-\cos k),
\end{equation}
where we defined $H(\xi) \equiv \sum_{x=1}^\infty
\sum_{y=-\infty}^\infty \FF{\zz}{ x, y}$, and where we used the
symmetry relation $\sum_{\ZZ\neq0} \widehat{F}^*(\zz|\ee_1|\ZZ) =
\sum_{\ZZ\neq0} \widehat{F}^*(\zz|-\ee_1|\ZZ)$. Consequently, the
determination of the CGF amounts to the calculation of the quantity
$H(\xi)$ and the first passage density $\widehat{F}_{ 1}$, which are
given in the SM~\cite{SM}.

\begin{figure}
\begin{center}
\includegraphics[width=.9\columnwidth,clip=true]{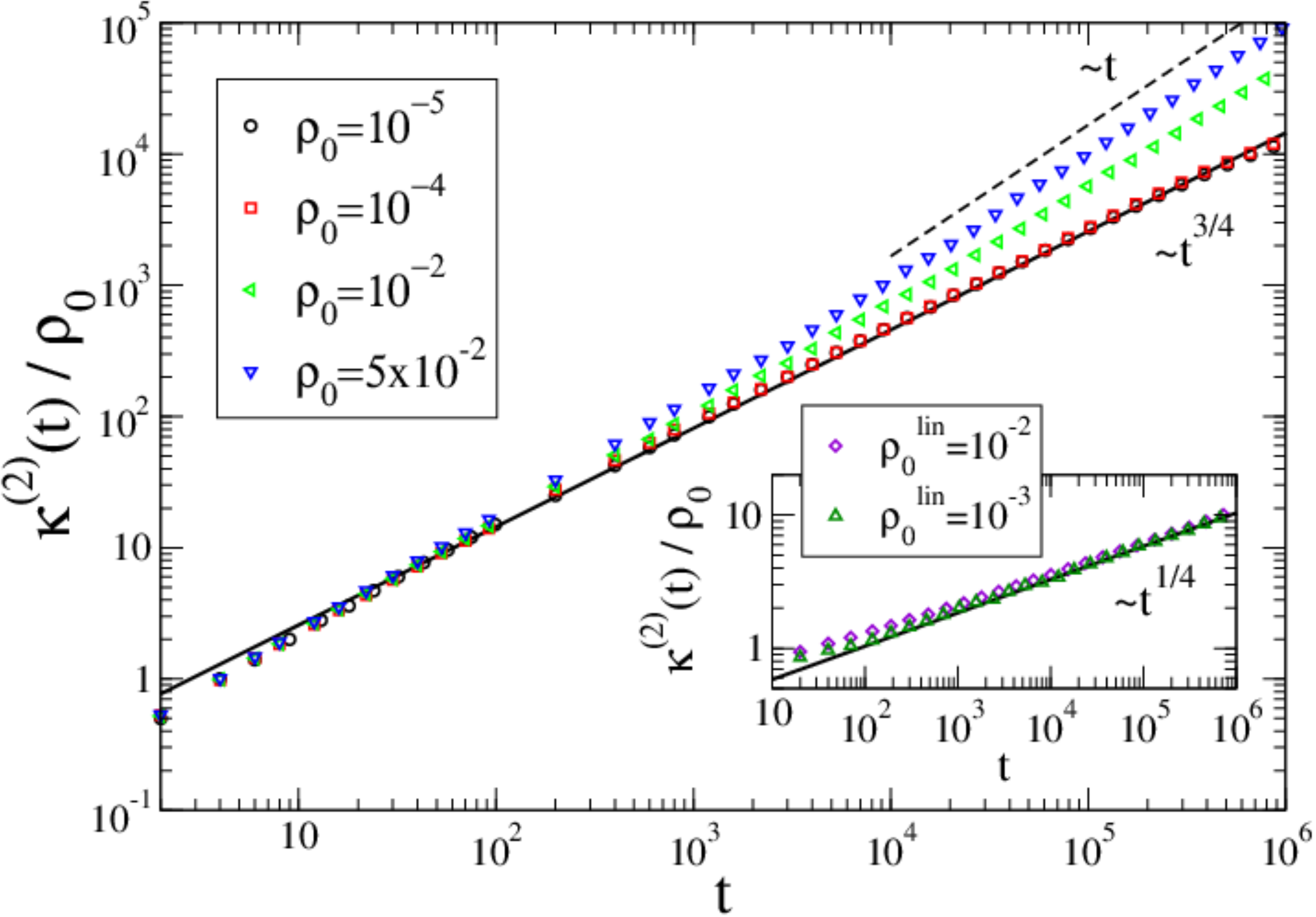}
\caption{(Color online) Variance rescaled by $\rho_0$ 
  computed from Monte-Carlo simulations of vacancy dynamics. The
  full line represents the analytical
  result~(\ref{result_uniform}). Inset: case
  where all vacancies are initially placed on the backbone (the full
  line represents Eq.~(\ref{result_backbone}.))}
\label{simu_2d_uniform}
\end{center}
\end{figure}

Expanding $\widehat{\Psi}(k;\xi)$ in powers of $k$ from
Eq.~(\ref{Psidef}), focusing on the large time limit $\xi\to 1^-$ and
using a Tauberian theorem \cite{Hughes1995} we then get the following
exact expression for the cumulants (see SM~\cite{SM} for details)
\begin{equation}
\lim_{\rho_0\to0} {\frac{{\kappa}^{(2n)}(t)}{\rho_0}} \underset{t\to\infty}{\sim}   \frac{1}{2^{5/4}\Gamma(7/4)}t^{3/4}.
\label{result_uniform}
\end{equation}
Several comments are in order. (i) First, one notices that all
cumulants of the same parity are equal. This indicates that the
probability distribution $P_t(X)$ is a Skellam distribution
\cite{Skellam1946}, originally defined as the p.d.f. associated with
the difference of two Poissonian random variables. In particular, the
rescaled position $X_t/t^{3/8}$ is asymptotically normally
distributed. (ii) The exponent $3/4$ is intermediate between a
``normal'' diffusion exponent, and the single-file
diffusion exponent $1/2$.  (iii) While the
analytical predictions of the cumulants are successfully compared to
results obtained from Monte Carlo simulations at intermediate times
(see Fig.~\ref{simu_2d_uniform}), at long times a crossover towards a
standard diffusive behavior is observed. 

We present below a
theoretical argument that accounts for this intriguing behavior.
The key point underlying this crossover is that the above
analytical results are derived by taking the limit $\rho_0\to 0$
before the long time limit.  In this case, to leading order, the TP
does not move before being reached by a given vacancy. In fact, the TP
actually diffuses due to its interactions with the other
vacancies. Because of this effective diffusion of the TP, each vacancy
experiences an additional symmetric jump probability in the $x$
direction, denoted by $D(\rho_0)$, even when they are on the teeth. 
Thus, in the reference frame of the TP, the vacancies are
allowed to jump from one tooth to another (with a
probability that vanishes as $\rho_0 \to 0$) and their motion is now
effectively two-dimensional. Qualitatively, we are facing a
two-dimensional situation, and regular diffusion is thus expected at
large times (see introduction). Quantitatively, the approach developed
previously can be extended to calculate the variance of $X_t$, in this
case where the vacancies perform 2D random walks, but can reach the TP
only from the backbone (see SM~\cite{SM}), yielding
\begin{equation}
\label{kappa_second}
\widehat{\kappa}^{(2)}(\xi) = -2 \rho_0  \frac{\Sigma(\xi,\rho_0)\left(\widehat{F}^*_{1}-\widehat{F}^*_{-1}-1\right)}{\left(\widehat{F}^*_{1}-1+\widehat{F}^*_{-1}\right)   \left( \widehat{F}^*_{1}+1-\widehat{F}^*_{-1}\right)},
\end{equation}
where we defined $\widehat{F}^*_{\pm1} \equiv
\widehat{F}^*(\zz|\ee_1|\ee_{\pm1};\xi,\rho_0)$ and
$\Sigma(\xi,\rho_0)\equiv\sum_{\ZZ\neq\zz}
\widehat{F}^*(\zz|\ee_1|\ZZ;\xi,\rho_0)$, with $F^*_t(\zz|\ee_1|\ZZ;
\rho_0)$ the probability for a vacancy to reach the origin for the
first time at time $t$ knowing that it was at site $\ee_1$ at time
$t-1$ and that it started from site $\ZZ$.
Relying on renewal-type equations, the conditional first-passage time
densities $\widehat{F}^*$ are related to the propagators of the
vacancies random walk, which are themselves evaluated with an
extension of the method presented in~\cite{Nieuwenhuizen2004} to treat
diffusion on inhomogeneous lattices (see SM~\cite{SM}). It is finally
found that
\begin{eqnarray}
  \widehat{F}^*_{-1} & \simeq & \frac{\sqrt{2}}{\pi} \sqrt{D(\rho_0)}\ln \frac{1}{ D(\rho_0)}- \frac{\pi \sqrt{ D(\rho_0)}}{\ln \frac{1}{1-\xi}} +\dots \label{def_F1}\\
  \widehat{F}^*_{1} & \simeq & 1- \frac{\pi \sqrt{ D(\rho_0)}}{\ln \frac{1}{1-\xi}} +\dots \label{def_Fm1}
\end{eqnarray}
where the symbol $\simeq$ stands for the long-time limit ($\xi \to 1$)
for a fixed value of $\rho_0$. As expected,
$\displaystyle\widehat{F}^*_{-1} \underset{\rho_0 \to 0}\to0$. Next,
$\Sigma(\xi,\rho_0)$ is conveniently estimated in SM~\cite{SM} by relying on the
continuous-space description given in~\cite{Arkhincheev1991}:
\begin{equation}
\label{dev_sigma}
\Sigma(\xi,\rho_0) \underset{\xi \to 1}{\propto} \frac{ \sqrt{D(\rho_0)}}{(1-\xi) \ln \frac{1}{1-\xi}}.
\end{equation} 
Plugging Eqs.~(\ref{def_F1}),~(\ref{def_Fm1}) and~(\ref{dev_sigma})
into~(\ref{kappa_second}), and using a Tauberian theorem, we finally
obtain
\begin{equation}
\label{variance_diff}
\lim_{t \to \infty} \frac{\kappa^{(2)}(t)}{t}   \underset{\rho_0 \to 0}{\propto} \rho_0  \sqrt{D(\rho_0)} \ln \frac{1}{D(\rho_0)}.
\end{equation}
This equation defines the effective diffusion coefficient
$D(\rho_0)=\lim_{t\to\infty} (\kappa^{(2)}(t)/{2t})$ self-consistently
when $\rho_0 \to 0$, and finally yields the following expression of
the variance in the ultimate regime:
\begin{equation}
\label{variance_diff2}
\lim_{t \to \infty} \frac{\kappa^{(2)}(t)}{t}    \underset{\rho_0 \to 0}{\propto} {\rho_0}^2 \left( \ln \frac{1}{\rho_0} \right)^2.
\end{equation}
These results thus show that the limits $\rho_0\to 0$ and $t\to\infty$
do not commute leading to an ultimate diffusive behavior \footnote{A
  similar non-inversion mechanism was found
  in~\cite{Benichou2013c}. However, in contrast to what is found here,
  this effect exists only when the TP experiences a non zero
  bias.}. However, due to a subtle coupling between EVIs and the
geometrical constraints involved in the comb geometry, this diffusive
regime displays a non analytical dependence on the vacancy density,
checked numerically in Fig.~4 of the SM~\cite{SM}.  This is markedly
different from the case of homogeneous lattices where a linear
behavior with $\rho_0$ is found~\cite{Brummelhuis1989a}.  In addition,
the comparison between Eqs.~(\ref{result_uniform})
and~(\ref{variance_diff2}) shows that the crossover time between the
two regimes behaves like $t_\times\sim (\rho_0\ln(\rho_0))^{-4}$,
which can be very large for dense systems. As a result, the
subdiffusive behavior of the first regime is long-lived and
potentially observable in real systems.  We now consider several
extensions of these results.

{\it Influence of the initial conditions.} The previous results were
obtained assuming that the vacancies were initially uniformly
distributed on the lattice. We now assume that they are initially
located only on the backbone, with a linear density
$\rho_0^{\rm lin}$ defined as the number of vacancies divided by the
length of the backbone.  Averaging over this initial 
distribution, which actually amounts to
restricting the sum over the initial points to the only backbone in
Eq.~(\ref{Psidef}), it is found that (see SM~\cite{SM})
\begin{equation}
{\kappa}^{(2n)}(t)  \underset{\rho_0^{\rm lin} \to 0}{\sim}  \frac{\rho_0^{\rm lin}}{2^{7/4}\Gamma(5/4)}t^{1/4}.
\label{result_backbone}
\end{equation}
Consequently, the time dependence of the cumulants is modified in a
dramatic way: the cumulants now grow
as $t^{1/4}$. This analytical prediction is successfully confronted to
numerical simulations (see inset of Fig.~\ref{simu_2d_uniform}). This
spectacular slowdown of the dynamics with respect to the uniform
initial conditions is compatible with Eq.~(\ref{result_uniform}), where
now $\rho_0$ strictly vanishes. Interestingly, in this case,
$t_\times\to\infty$, so that there is no cross-over to a diffusive
regime.

\emph{$d$-dimensional comb.}-- The previous results can also be
generalized to the important case of a $d$-dimensional comb
$\mathbf{C}_d$, widely studied in the
literature~\cite{Ben-avraham,Cassi1992,Bertacchi2003}, which is
defined recursively: starting from $\mathbf{C}_1$ (a one-dimensional
lattice), $\mathbf{C}_d$ is obtained from $\mathbf{C}_{d-1}$ by
attaching at each point a two-way infinite path (see figure in
SM~\cite{SM}).  It is found that for uniform initial conditions, the
even cumulants all behave like
\begin{equation}
\lim_{\rho_0\to0}\frac{\kappa^{(2n)}(t)}{\rho_0} \underset{t \to \infty}{\propto}  t^{1-\frac{1}{2^d}},
\end{equation}
and eventually cross-over to a diffusive linear in time regime for $d
\geq 2$.  Note that in the case of the $d=1$ comb, single-file
subdiffusion $ \kappa^{(2n)}(t)\sim\sqrt{t}$ is recovered.

Finally, reminding that single-file diffusion has been shown to be a
realization of a fractional Brownian motion with Hurst exponent
$1/4$~\cite{Landim2000}, we conjecture
that tracer diffusion in a crowded $d$-comb is more generally a
realization of a fractional Brownian motion of Hurst exponents
$H=(2^d-1)/2^{d+1}$.

{\it Case of a TP visiting the teeth}-- We finally come back to the
original problem of a tracer on a crowded 2-comb, where the TP is
identical to the bath particles and thus allowed to visit the teeth of
the comb. The displacement of the TP along the backbone can be
analyzed in a mean-field description that decouples the motion of the
TP in a tooth from the dynamics of other bath particles as a
Continuous Time Random Walk, whose waiting time distribution $\psi(t)$
describes the time the TP spends on a given tooth of the crowded comb.
\begin{figure}
\begin{center}
\includegraphics[width=1.\columnwidth,clip=true]{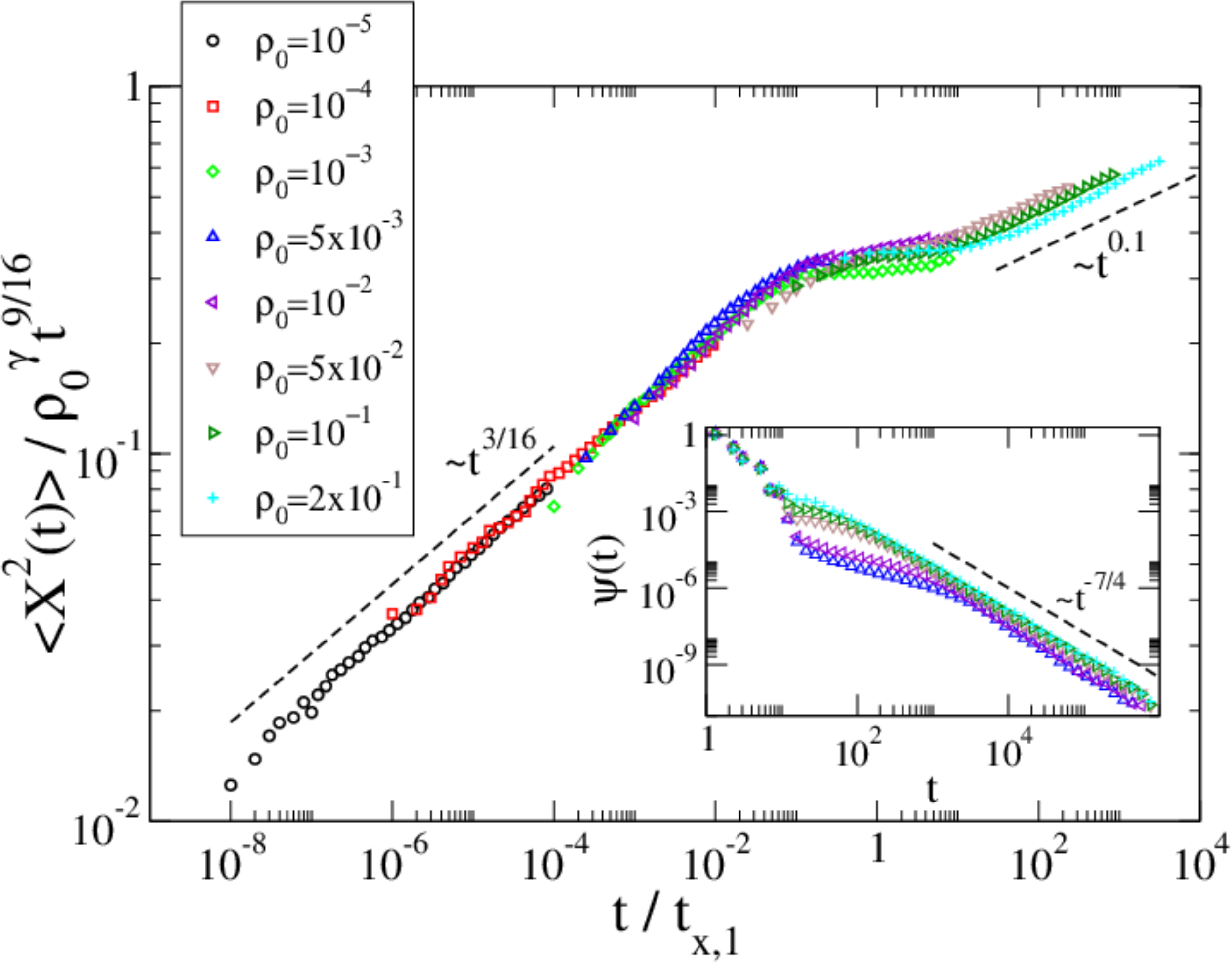}
\caption{Variance of the TP visiting the teeth, measured for different
  values of $\rho_0$. The collapse of the curves is obtained by
  rescaling time by the crossover time $t_{\times,1}=1/\rho_0^2$,
  predicted by our analytical approach, and the variance by
  $\rho_0^\gamma$ with $\gamma\simeq 0.62$, numerically obtained. The
  plateau at intermediate times confirms the prediction of
  Eq.~(\ref{CTRW}).  For $\rho_0\le 10^{-3}$ the reported results are
  obtained via numerical simulations of vacancy dynamics, while for
  $\rho_0> 10^{-3}$ via particle dynamics (see SM~\cite{SM}).}
\label{simu}
\end{center}
\end{figure} 
Noting that the motion of the TP along a tooth is close to a
single-file motion, we expect that the transverse MSD behaves like
$\langle y^2(t)\rangle \propto \sqrt{\rho_0^2 t}$~\cite{Arratia1983},
in the dense limit. In turn, this leads to two different regimes for
$\psi(t)$: for $t\ll t_{\times,1}\equiv 1/\rho_0^2$, $\langle
y^2(t)\rangle \ll 1$, the TP has not had time to explore a tooth
because of the other crowding bath particles of the tooth, and the
mean time spent on the tooth is finite; for $t\gg t_{\times,1}$,
$\psi(t)\propto 1/t^\mu$, with $\mu=7/4$, as obtained in~\cite{Krug1997,Molchan1999} 
and checked numerically (see inset of Fig~\ref{simu}).

The MSD $\langle X^2\rangle$ of the TP along the backbone is then
related to the MSD $\kappa^{(2)}$ of the TP restricted to the backbone
by using the standard Montroll Weiss relation~\cite{Hughes1995}:
\begin{equation}
\widehat{ \langle X^2\rangle}(\xi)=\frac{1-\widehat{\psi}(\xi)}{1-\xi}\widehat{\kappa^{(2)}}(\widehat{\psi}(\xi)).
\end{equation}
Combining the two temporal regimes of $\kappa^{(2)}(t)$ determined
previously with the two regimes of $\psi(t)$, we finally obtain that
$\langle X^2(t)\rangle$ displays three non trivial regimes :
 \begin{equation}
 \label{CTRW}
\langle X^2(t)\rangle\propto 
\begin{cases}
      t^{3/4}& \text{if $t\ll t_{\times,1}$ }, \\
       t^{3/4(\mu-1)} = t^{9/16}&\text{if $t_{\times,1} \ll t\ll t_{\times,2}$},\\
       t^{\mu-1} = t^{3/4}&\text{if  $t\gg t_{\times,2}$}
\end{cases}
\end{equation}
where $ t_{\times,2}$ is a second crossover time (whose explicit
dependency on $\rho_0$ is not provided by our approach). The
comparison with numerical simulations shown in Fig.~\ref{simu} reveals
that: (i) three temporal regimes with expected crossover time
$t_{\times,1}$ are indeed observed; (ii) the exponents of the two
first are in good agreement with our analytical
prediction~(\ref{CTRW}); (iii) the increase of $\langle X^2(t)\rangle$
observed in the last regime is in qualitative agreement with
~(\ref{CTRW}) but the quantitative determination of the corresponding
exponent would require more extensive simulations.  Remarkably, it is
found that in all regimes, the dynamics of the TP along the backbone
is faster than in the absence of bath particles, where $\langle
X^2(t)\rangle\sim t^{1/2}$. In other words, the motion of the TP is
accelerated along the backbone by EVIs.  This
surprising behavior results from two competing effects quantified by
our approach: hard-core interactions hinder the motion of the TP along
the backbone but in the same time reduce the time lost by the TP in
the teeth.

\onecolumngrid

\newpage

\begin{center}

\Large{\textbf{Supplemental Material}}

\end{center}

\section{Evolution rules of the vacancies}
\label{evol_rules}

In the main text, we first studied the case where the tracer particle
(TP) is \emph{constrained to move on the backbone of the comb}. The
evolution rules of the particles where given in Fig. 1: (i) the TP
jumps on each of the neighboring sites of the backbone with
probability $1/2$; (ii) if a bath particle is located on a tooth of
the comb, it jumps on each of the neighboring sites with probability
$1/2$; (iii) if a bath particle is located on the backbone of the
structure, it jumps on each of the neighboring sites with probability
$1/4$. The dynamics is constrained with hardcore interactions, which
means that there is at most one particle per site.

\begin{figure}
\begin{center}
\includegraphics[width=9cm]{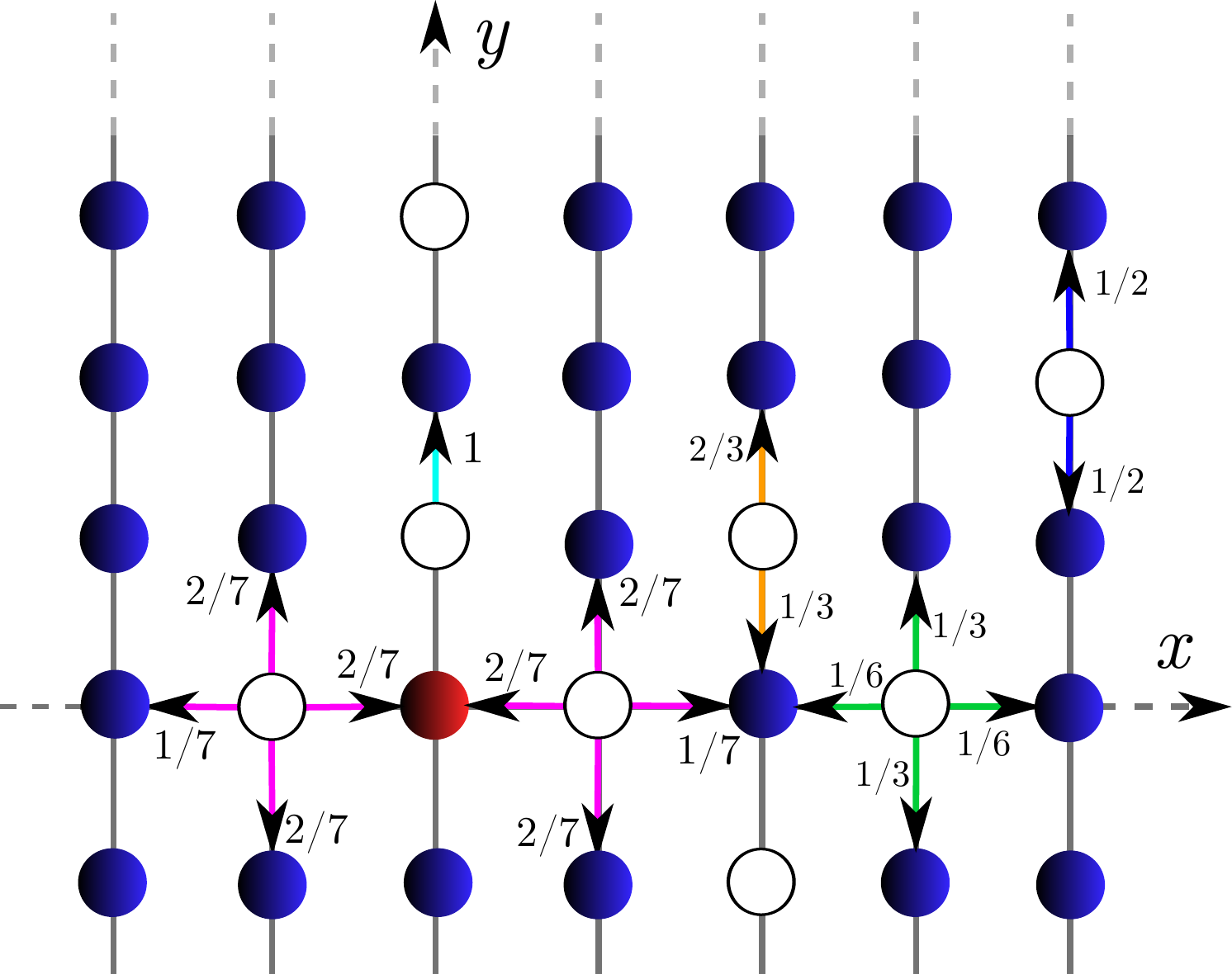}
\caption{Evolution rules of a vacancy on the two-dimensional comb.}
\label{fig:2_comb_vac_dynamics_sym}
\end{center}
\end{figure}

In the high-density limit, there are very few vacancies on the
lattice, and it is more convenient to describe their dynamics instead
of describing the dynamics of all the particles. The events where two
vacancies are on neighboring sites are of order
$\mathcal{O}({\rho_0}^2)$ (where $\rho_0$ is the density of vacancies
on the lattice). As we focus on the results at leading order in
$\rho_0$, we do not take these events into account. We adopt a
discrete-time evolution, and assume that at each time step, each
vacancy exchanges its position with one of the neighboring
particles. Depending on the position $\ZZ$ of the vacancy (see
Fig. \ref{fig:2_comb_vac_dynamics_sym}), its jump probabilities are
defined as follows:
\begin{itemize}
\item if the vacancy is on a tooth of the comb but not adjacent to the
  backbone, it exchanges its position with the same probability with
  each of the neighboring particles.
\item if the vacancy is at $\ZZ$, adjacent to the backbone but not to
  the TP, the particle located at $\ZZ-\ee_2$ (resp. $\ZZ+\ee_2$) has
  a probability $1/4$ (resp. $1/2$) to exchange its position with that
  of the vacancy. Consequently, the vacancy has a probability
  proportional to $1/4$ (resp. proportional to $1/2$) to jump in
  direction $-\ee_2$ (resp. $\ee_2$), so that
  \begin{eqnarray}
p(  \ZZ-\ee_2|\ZZ) & = & \mathcal{Z} \times \frac{1}{4}, \\
p(  \ZZ+\ee_2|\ZZ) & = & \mathcal{Z} \times \frac{1}{2},
\end{eqnarray}
  where $p(\rr|\rr')$ is the probability for a vacancy to jump from site $\rr'$ to site $\rr$ in a single step, and where $\mathcal{Z}$ is a normalization constant. With the normalization condition $p(\ZZ \rightarrow \ZZ-\ee_2)+p(\ZZ \rightarrow \ZZ+\ee_2)=1$, we find $Z=4/3$, and finally
    \begin{eqnarray}
p(  \ZZ-\ee_2|\ZZ) & = & \frac{1}{3}, \\
p(  \ZZ+\ee_2|\ZZ) & = &  \frac{2}{3}. 
\end{eqnarray}
    \item if the vacancy is on the backbone but not adjacent to the TP, we can adapt the previous arguments to write
    \begin{eqnarray}
p(  \ZZ\pm\ee_1|\ZZ) & = & \frac{1}{6}, \\
p(  \ZZ\pm\ee_2|\ZZ) & = & \frac{1}{3}. 
\end{eqnarray}
\item if the vacancy is on the backbone and at the right of the TP (this is easily extended to the case where the vacancy is at the left of the TP), the same arguments lead to
\begin{eqnarray}
p(  \ZZ-\ee_1 |\ZZ) & = & \frac{2}{7}, \\
p(  \ZZ+\ee_1 |\ZZ) & = & \frac{1}{7}, \\
p(  \ZZ\pm\ee_2 |\ZZ) & = & \frac{2}{7}. 
\end{eqnarray}

\item if the vacancy is on a tooth adjacent to the TP, it steps away from the TP with probability $1$ as the TP is constrained to remain on the backbone.

\end{itemize}

These evolution rules are summarized in Fig. \ref{fig:2_comb_vac_dynamics_sym}.

\section{First regime}

In the main text, we established the following expression for the generating function associated with the cumulant generating function of the position of the TP:
\begin{equation}
\label{Psidef2}
{\widehat{\Psi}(k;\xi)} \underset{\rho_0 \to 0}{\sim} - 2 \rho_0 \frac{ H(\xi) (1-\cos k)}{(1-\xi)(1+\widehat{F}_1)} ,
\end{equation}
where we use the simplified notation $\widehat{F}_{ 1} \equiv \widehat{F}(\zz|\ee_{ 1};\xi)$, and where   $\widehat{F}(\zz|\rr;\xi)$ is the generating function associated with $F_t(\zz|\rr)$, defined as the probability for a vacancy to reach the origin for the first time at time $t$ knowing that it started from site $\rr$. We also introduced the quantity $H (\xi) \equiv \sum_{x=1}^\infty \sum_{y=-\infty}^\infty  \FF{\zz}{ x, y}$. The determination of the cumulant generating function then relies on the determination of $H(\xi)$ and $\widehat{F}_1$. Detailed calculations of these quantities are given below.

\subsection{Calculation of $H(\xi)$}
\label{sec:H}

The quantity $H(\xi)$ is defined as
\begin{equation}
H(\xi)  =  \sum_{x=1}^\infty \sum_{y=-\infty}^\infty  \FF{0,0}{ x,y}.
\end{equation}
The first-passage densities  $\FF{0,0}{ x,y}$ are related to the random walk of a vacancy on a lattice with specific and non-uniform transition rates, and represented in Fig. \ref{fig:2_comb_vac_dynamics_sym}. 

These first-passage densities can be calculated using the following fundamental relation, resulting from the tree-like structure of the two-dimensional comb, and valid for any site $\rr'$ belonging to the path from $\rr$ to $\rr_0$ \cite{Woess2000}:
\begin{equation}
\label{eq:prop_tree}
\widehat{F}(\rr|\rr_0;\xi)=\widehat{F}(\rr|\rr';\xi)\widehat{F}(\rr'|\rr_0;\xi).
\end{equation}
To compute $\FF{0,0}{ x,y}$ ($x>0$), we consider separately the situations where $y=0$ and $y\neq 0$, and obtain:
\begin{itemize}
	\item if $y \neq 0$, the path from $( x,y)$ to $(0,0)$ is decomposed as follows:
	\begin{equation}
	( x,y) \rightarrow ( x,\mathrm{sgn}(y)) \rightarrow ( x,0) \rightarrow (1 ,0) \rightarrow (0,0)
	\end{equation}
	so that generating function of the first-passage density is
	\begin{equation}
	\FF{0,0}{ x, y}  =  \FF{0,0}{1,0} \FF{1,0}{ x,0}  \FF{ x,0}{ x,\mathrm{sgn}(y)}\FF{ x,\mathrm{sgn}(y)}{ x,y}.\\
	\end{equation}
	We introduce the quantities 
	\begin{eqnarray}
	f_{2,1}(\xi) & = & \FF{1,0}{2,0}, \label{fff1} \\
	f_{2,2}(\xi) & = & \FF{1,0}{1,1},  \label{fff2} \\
	f_{1,1}(\xi) & = & \FF{1,1}{1,2}. 
	\end{eqnarray}
	The last quantity is relative to the random walk of a vacancy on a tooth far from the backbone, i.e. on a one-dimensional structure. It is then given by \cite{Hughes1995}:
	\begin{equation}
	f_{1,1}(\xi) = \frac{1-\sqrt{1-\xi^2}}{\xi}.
        \label{f11}
	\end{equation}
	Finally, using these relations and the fundamental equation (\ref{eq:prop_tree}) once again, we get
	\begin{equation}
	\FF{0,0}{ x, y}	= \widehat{F}_1  \left[ f_{2,1}(\xi)  \right]^{|x|-1} f_{2,2}(\xi) \left[ f_{1,1}(\xi)  \right]^{|y|-1}.
	\end{equation}
	\item  if $y=0$, the path is simply decomposed as:
	\begin{equation}
	 ( x,0) \rightarrow (1 ,0) \rightarrow (0,0)
	\end{equation}
	and we find
	\begin{eqnarray}
	\FF{0,0}{ x, 0}  &=&  \FF{0,0}{1,0} \FF{1,0}{ x,0} \\
	 &=& \widehat{F}_1 \left[ f_{2,1}(\xi)  \right]^{|x|-1} .
	\end{eqnarray}
\end{itemize}

We now go back to the calculation of $H(\xi)$,
\begin{eqnarray}
H(\xi) & = & \sum_{x=1}^\infty \sum_{y=-\infty}^\infty \FF{0,0}{ x,y} \\
& = & \sum_{x=1}^\infty  \left[  \FF{0,0}{ x, 0}  +2 \sum_{y=1}^\infty \FF{0,0}{ x , y}  \right] \\
& = & \widehat{F}_1 \sum_{x=1}^\infty f_{2,1}(\xi)^{x-1} \left[ 1 + 2\sum_{y=1}^\infty f_{2,2}(\xi) f_{1,1}(\xi)^{y-1}    \right] \\
&=& \frac{ \widehat{F}_1}{1-f_{2,1}(\xi)} \left( 1+2 \frac{f_{2,2}(\xi)}{1-f_{1,1}(\xi)} \right)\label{exp_H}   .
\end{eqnarray}
%We will then write
%\begin{equation}
%\label{eq:hnu_simple_2-comb}
%h_\epsilon \equiv \widehat{F}_\epsilon h(\xi)
%\end{equation}
%with
%\begin{equation}
%\label{eq:def_h_2-comb}
%h(\xi) \equiv  \frac{ 1}{1-f_{2,1}(\xi)} \left( 1+2 \frac{f_{2,2}(\xi)}{1-f_{1,1}(\xi)} \right) .
%\end{equation}

We then calculate separately $f_{2,2}(\xi)$ and $f_{2,1}(\xi)$, defined in Eqs.~(\ref{fff1}) and~(\ref{fff2}), which are relative to the random walk of a vacancy away from the tracer.\\
% , and defined respectively by
% \begin{eqnarray}
% f_{2,2}(\xi)&=&\widehat{F}(\ee_1|\ee_1+\ee_2;\xi) \\
% f_{2,1}(\xi)&=&\widehat{F}(\ee_1|2\ee_1;\xi) 
% \end{eqnarray}

In what follows, we will use  the general relation
\begin{equation}
\label{eq:partition_Fstar}
\widehat{F}(\rr|\rr_0;\xi)=\xi \sum_{\boldsymbol{w}}\widehat{F}(\rr|\boldsymbol{w};\xi)p(\boldsymbol{w}|\rr_0),
\end{equation}
where $p(\boldsymbol{w}|\rr_0)$ is the probability to jump from $\rr_0$ to $\boldsymbol{w}$ in a single step, and where the sum runs over all lattice sites. This relation is obtained by partitioning over the first step of the walk.

%\begin{enumerate}
	\subsubsection{ Calculation of $f_{2,2}(\xi)$}

	We use the general relation (\ref{eq:partition_Fstar}), and write
	\begin{eqnarray}
	f_{2,2}(\xi)&=&\xi \sum_{\boldsymbol{w}}\widehat{F}(\ee_1|\boldsymbol{w};\xi)p(\boldsymbol{w}|\ee_1+\ee_2) \\
			&=&\xi \left[\widehat{F}(\ee_1|\ee_1;\xi)p(\ee_1|\ee_1+\ee_2) + \widehat{F}(\ee_1|\ee_1+2\ee_2;\xi)p(\ee_1+2\ee_2|\ee_1+\ee_2) \right]
	\end{eqnarray}
	Because of the tree structure of the lattice, one writes
		\begin{eqnarray}
		\widehat{F}(\ee_1|\ee_1;\xi)&=&1 \\
		\widehat{F}(\ee_1|\ee_1+2\ee_2;\xi)&=&\widehat{F}(\ee_1|\ee_1+\ee_2;\xi)\widehat{F}(\ee_1+\ee_2|\ee_1+2\ee_2;\xi)=f_{2,2}(\xi)f_{1,1}(\xi).
		\end{eqnarray}
	Finally, we get the following equation for $f_{2,2}(\xi)$:
	\begin{equation}
	f_{2,2}(\xi)= \xi \left[\frac{1}{3} +\frac{2}{3} f_{1,1}(\xi) f_{2,2}(\xi)   \right].
	\end{equation}
	Solving it and recalling the expression of $f_{1,1}(\xi)$ (Eq.~(\ref{f11})), we obtain
	\begin{equation}
	f_{2,2}(\xi)=\frac{\xi}{1+2\sqrt{1-\xi^2}}.
	\end{equation}
	
	\subsubsection{ Calculation of $f_{2,1}(\xi)$}
	
	Using again Eq. (\ref{eq:partition_Fstar}), we write
	\begin{eqnarray}
	f_{2,1}(\xi)  &=& \xi \sum_{\boldsymbol{w}}\widehat{F}(\ee_1|\boldsymbol{w};\xi)p(\boldsymbol{w}|2\ee_1) \\
	&=& \xi\left[\FF{\ee_1}{\ee_1} p(\ee_1|2\ee_1) +\FF{\ee_1}{3 \ee_1} p(3 \ee_1|2\ee_1) +2 \FF{\ee_1}{2\ee_1+\ee_2} p(2\ee_1+\ee_2|2\ee_1)   \right]\\
	&=& \xi \left[ \frac{1}{6} + \frac{1}{6}f_{2,1}(\xi)^2+\frac{2}{3} f_{2,2}(\xi)f_{2,1}(\xi)    \right].
	\end{eqnarray}
	$f_{2,1}(\xi)$ is then the solution of a second-order equation. Choosing the solution fulfilling the condition $f_{2,1}(0)=0$, we get
	\begin{equation}
		f_{2,1}(\xi)=\frac{3}{\xi}-2f_{2,2}(\xi)-\sqrt{\left(\frac{3}{\xi}-2 f_{2,2}(\xi)   \right)^2-1}.
	\end{equation}

%\end{enumerate}

\subsection{Calculation of $\widehat{F}_{ 1}$}

%\textcolor{red}{a specifier au cas sym}

In order to calculate the generating function $\widehat{F}_1= \widehat{F}(\zz|\epu;\xi)$, we consider the first passage density $ {F}_t(\zz|\epu)$ partitioned over the first time step
\begin{equation}
\Ft{\zz}{\ee_1} = \frac{2}{7} \delta_{t,1} + \frac{1}{7} \sum_{t'=1}^t {F}_{t'-1}( \ee_1 |2\ee_1){F}_{t-t'}( \zz |\ee_1) + 2\times \frac{2}{7} \sum_{t'=1}^t {F}_{t'-1}( \ee_1|\ee_1+\ee_2){F}_{t-t'}( \ee |\ee_1).
\end{equation}
The first term of right-hand side corresponds to the situation where the vacancy  jumps onto the origin of the lattice at the first step (i.e. in direction $-\ee_1$). The second (resp. third) term corresponds to the situation where the first step of the vacancy is in direction $+\ee_1$ (resp. $ \pm \ee_2$). In terms of generating functions, we get
\begin{eqnarray}
\widehat{F}_1&=&\frac{\frac{2}{7} \xi}{1-  \frac{1}{7}\xi\FF{\ee_1}{2\ee_1}-   \frac{4}{7}\xi\FF{\ee_1}{\ee_1+\ee_2}}\\
&=& \frac{\frac{2}{7} \xi}{1-\frac{1}{7}\xi f_{2,1}(\xi)-\frac{4}{7}\xi f_{2,2}(\xi)} \label{exp_F1}.
\end{eqnarray}
%Similarly, we get an expression for $\widehat{F}_{-1}$,
%\begin{equation}
%\label{eq:def_Fm1_2-comb}
%\widehat{F}_{-1}= \frac{r_1 \xi}{1-r_{-1}\xi f_{2,1}(\xi)-(1-r_1-r_{-1})\xi f{2,2}(\xi)},
%\end{equation}
%with
%\begin{eqnarray}
%r_{1} & = & \frac{p_{-1}}{p_{-1}+5/4}, \\
%r_{-1} & = & \frac{1/4}{p_{-1}+5/4}. 
%\end{eqnarray}

%Using Eqs. \ref{dev_f_1}, \ref{dev_f_2} and \ref{dev_g_2}, it is straightforward to establish the following expansions:
%\begin{equation}
%\widehat{F}_{\pm1}  \underset{\xi\to 1^-}{=}  1-\frac{1}{2^{1/4}p_{\pm1}} (1-\xi)^{1/4} +\mathcal{O}(\sqrt{1-\xi}).
%\end{equation}

\subsection{Long-time expansion of the cumulants}
\label{sec:cumulants}

Recalling the expression of $\widehat{\Psi}(k;\xi)$ from Eq. (\ref{Psidef2}):
\begin{equation}
{\widehat{\Psi}(k;\xi)} \underset{\rho_0 \to 0}{\sim} - 2 \rho_0 \frac{ H(\xi) (1-\cos k)}{(1-\xi)(1+\widehat{F}_1)} ,
\end{equation}
we write its expansion in powers of $k$ and obtain
\begin{equation}
\label{devPsi1}
{\widehat{\Psi}(k;\xi)} \underset{\rho_0 \to 0}{\sim} - 2 \rho_0 \frac{ H(\xi) }{(1-\xi)(1+\widehat{F}_1)} \sum_{n=1}^\infty (-1)^{n+1}\frac{k^{2n}}{(2n)!}.
\end{equation}
From the definition of the cumulant generating function $\widehat{\Psi}(k;\xi)$, the coefficients of its expansion in powers of $k$ are related to the cumulants of $X_t$ as follows:
\begin{equation}
\label{devPsi2}
\widehat{\Psi}(k;\xi) = \sum_{n=0}^\infty \frac{(\ii k)^n}{n!} \widehat{\kappa}^{(n)}(\xi).
\end{equation}
Identifying the expansions (\ref{devPsi1}) and (\ref{devPsi2}), we obtain that all the even cumulants are equal and have the following expression
\begin{equation}
\widehat{\kappa}^{(2n)} (\xi) \underset{\rho_0 \to 0}{\sim} \frac{2 \rho_0 H(\xi)}{(1-\xi)(1+\widehat{F}_1)}.
\end{equation}
We also verify that, for symmetry reasons, the odd cumulants are null. Using the explicit expressions of $H(\xi)$, Eq.~(\ref{exp_H}), and $\widehat{F}_1$, Eq.~(\ref{exp_F1}), we obtain the expansion of the even cumulants in the long-time limit ($\xi \to 1$):
\begin{equation}
\lim_{\rho_0 \to 0} \frac{\widehat{\kappa}^{(2n)}(\xi)}{\rho_0}  \underset{\xi \to 1}{ \sim}  \frac{1}{2^{5/4} (1-\xi)^{7/4}}. 
\end{equation}
Using a Tauberian theorem, we retrieve the long-time limit of these cumulants, corresponding to Eq.~(5) from the main text,
\begin{equation}
\lim_{\rho_0 \to 0} \frac{{\kappa}^{(2n)}(t)}{\rho_0}  \underset{t \to \infty}{ \sim}  \frac{1}{2^{5/4} \Gamma(7/4)} t^{3/4}. 
\end{equation}

\subsection{Influence of the initial conditions}

In the previous sections, we considered the situation where the vacancies are initially uniformly distributed on the lattice.  The situation where the vacancies are initially located on the backbone of the structure can be studied with the formalism presented above. It is straightforward to show that, with this initial condition, the cumulant generating function becomes
\begin{equation}
%\label{Psidef}
{\widehat{\Psi}(k;\xi)} \underset{\rho_0 \to 0}{\sim} - 2 \rho_0 \frac{ H'(\xi) (1-\cos k)}{(1-\xi)(1+\widehat{F}_1)} ,
\end{equation}
where
\begin{equation}
\label{exp_Hp}
H'(\xi)  \equiv \sum_{x=1}^\infty  \FF{\zz}{ x, y=0}.
\end{equation}
The expression of the even cumulants is deduced as in section \ref{sec:cumulants}, and we obtain
\begin{equation}
\widehat{\kappa}^{(2n)}(\xi)\underset{\rho_0\to0}{\sim} \frac{2\rho_0\widehat{F}_1}{(1-\xi)(1+\widehat{F}_1)[1-f_{2,1}(\xi)]}.
\end{equation}
Finally, recalling the expressions of $\widehat{F}_1$ (Eq. (\ref{exp_F1})) and $H'(\xi)$ (Eq. (\ref{exp_Hp})) in terms of the first-passage time densities computed in section \ref{sec:H}, we obtain the long-time limit ($\xi\to1$) of the even cumulants:
\begin{equation}
\lim_{\rho_0 \to 0} \frac{\widehat{\kappa}^{(2n)}(\xi)}{\rho_0}  \underset{\xi \to 1}{ \sim}  \frac{1}{2^{7/4} (1-\xi)^{5/4}}. 
\end{equation}
Using a Tauberian theorem, we retrieve the long-time limit of these cumulants:
\begin{equation}
\lim_{\rho_0 \to 0} \frac{{\kappa}^{(2n)}(t)}{\rho_0}  \underset{t \to \infty}{ \sim}  \frac{1}{2^{7/4} \Gamma(5/4)} t^{1/4}.
\end{equation}

\subsection{Definition of a $d$-dimensional comb}

\begin{figure}
\begin{center}
\includegraphics[width=8cm]{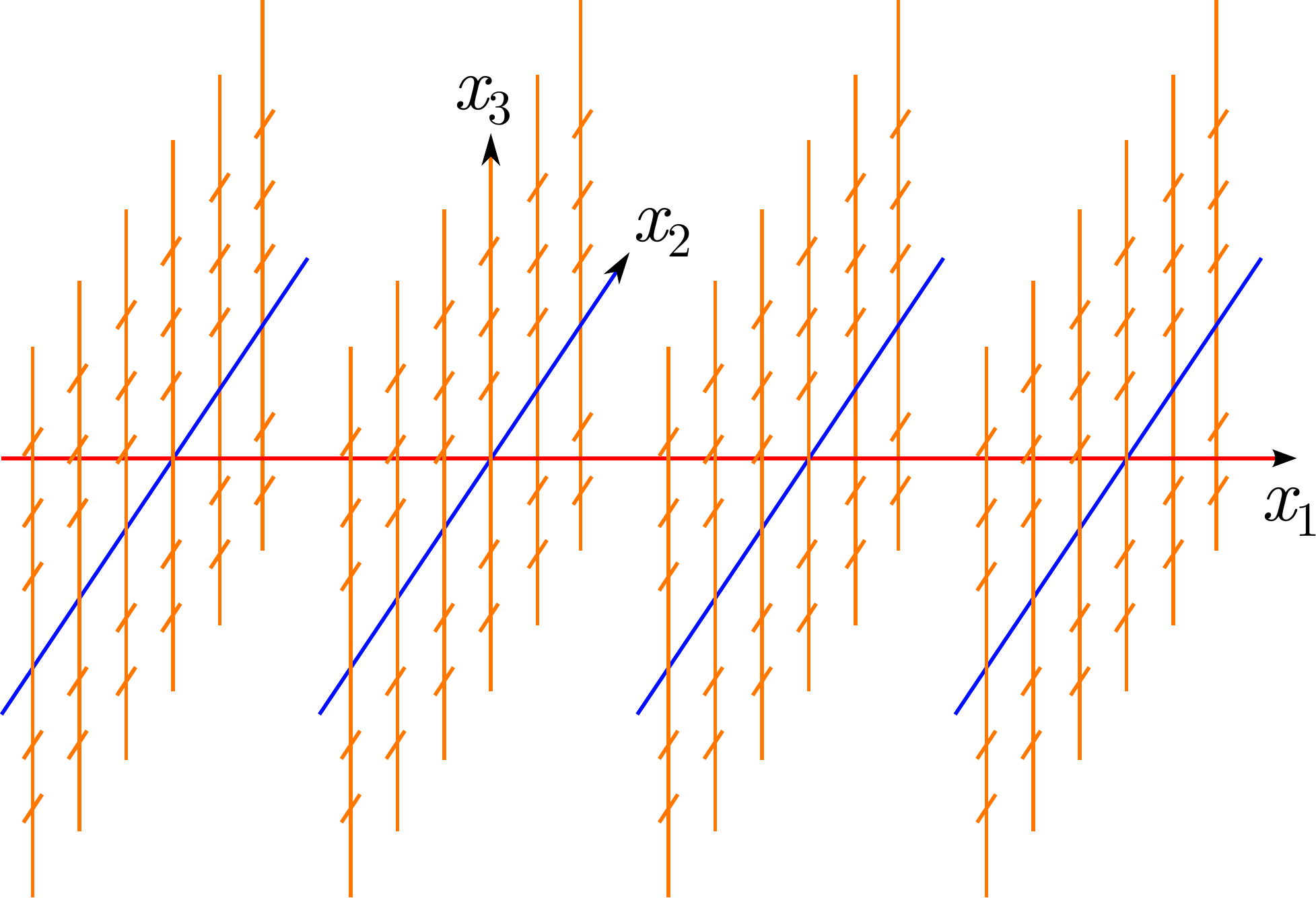}
\caption{Representation of the three-dimensional comb lattice.}
\label{3comb}
\end{center}
\end{figure}

In the main text, we present the expression of the variance of the
position of the TP on a generalized $d$-dimensional comb. This
structure is defined recursively as follows: starting from
$\mathbf{C}_1$ (a one-dimensional lattice), $\mathbf{C}_d$ is obtained
from $\mathbf{C}_{d-1}$ by attaching at each point a two-way infinite
path. We represent in Fig.~\ref{3comb} the three-dimensional comb.

\section{Second regime}

\subsection{Introduction}

\begin{figure}
\begin{center}
\includegraphics[width=8cm]{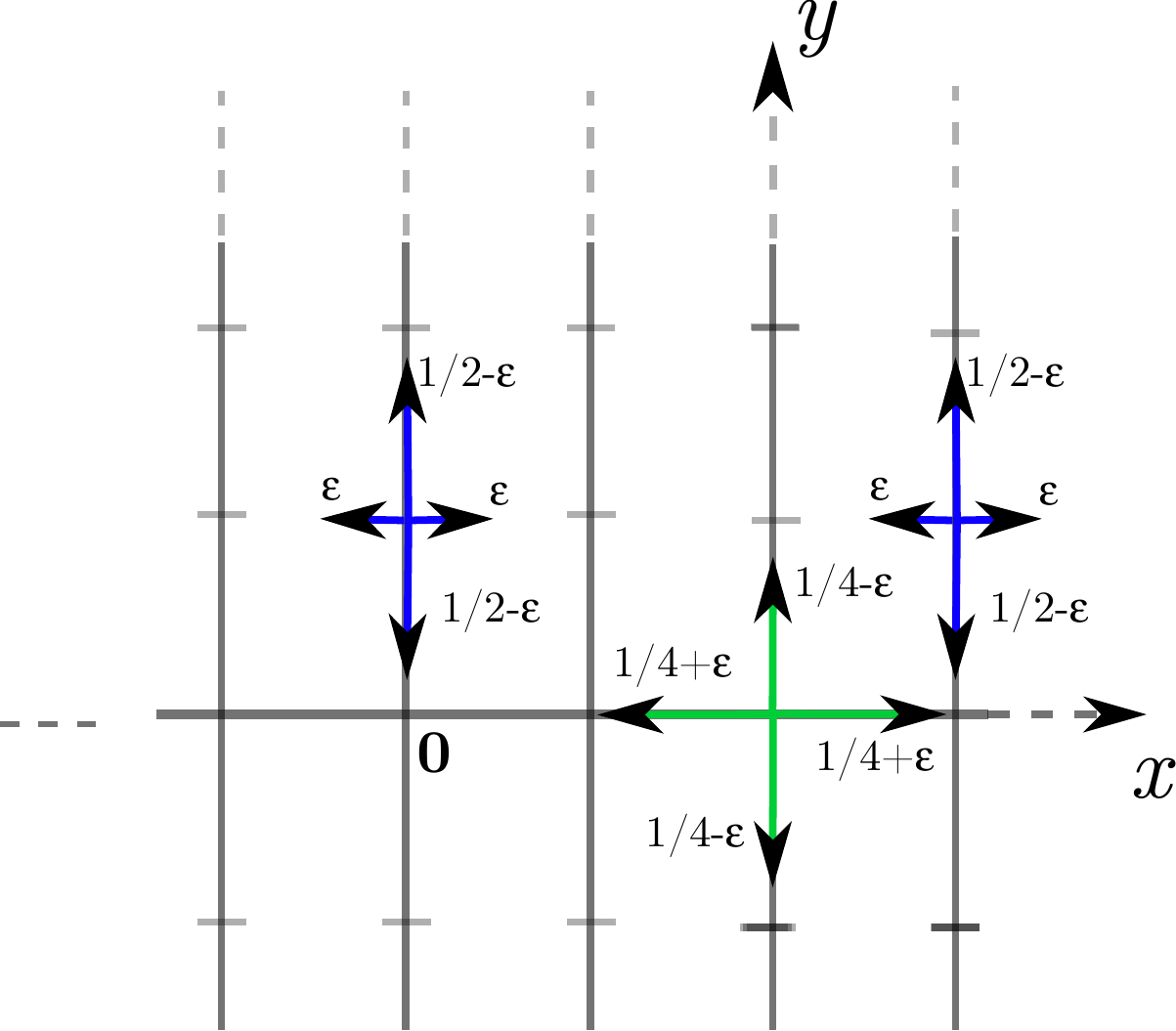}
\caption{Simplified evolution rules of the vacancies, when they experience an additional jump probability $\ep\equiv D(\rho_0)$.}
%{\emph{Left:} Simplified dynamics of the vacancies in the second regime, with a perturbation at the origin of the lattice. This lattice is denoted as $\Omega^\dagger$. \emph{Right:} Unperturbed random walk of a vacancy with simplified dynamics. This lattice is denoted as $\Omega$.}
\label{simplified_dynamics}
\end{center}
\end{figure}

The key point underlying the crossover to a second regime is that the
previous analytical results are derived by taking the limit $\rho_0\to
0$ before the long time limit. In this case, to leading order, the TP
does not move before being reached by a given vacancy. In fact, the TP
actually diffuses due to its interactions with the other vacancies.
Due to the effective diffusion of the TP, each vacancy is then assumed
to have an additional jump probability in the $x$ direction, denoted
by $D(\rho_0)$, and which vanishes when $\rho_0 \to 0$. We will write
for simplicity $\ep \equiv D(\rho_0)$. The vacancies then perform
two-dimensional random walks. We consider simplified evolution rules
of the vacancies, which are expected to give a correct qualitative
description of the system. The evolution rules for the vacancies are
represented in Fig. \ref{simplified_dynamics}.

\subsection{Single-vacancy problem}

As in the previous calculation, we first consider the situation where
there is only one vacancy on the lattice. We denote by
$P_t^{(1)}(\boldsymbol{X}|\boldsymbol{Y}_0;\ep)$ the probability to find
the TP at position $\boldsymbol{X}$ at time $t$ knowing that the
vacancy started from site $\boldsymbol{Y}_0$. This probability can be
written in terms of the first-passage time densities $F_t(\zz|\rr;\ep)$
(probability for a vacancy to reach the origin for the first time at
time $t$ starting from $\rr$) and the conditional first-passage
densities $F^*_t(\zz|\ee_\mu|\rr;\ep)$ (probability for a vacancy to reach
the origin for the first time at time $t$ starting from $\rr$ and
being at site $\ee_\mu$ at time $t-1$). Summing over the number of steps
$p$ taken by the TP, over the directions $\nu_1,\dots,\nu_p$ of these
$p$ steps and over the length of the time intervals $m_j$ elapsed
between the $(j-1)$-th and $j$-th step, we get
\begin{eqnarray}
P^{(1)}_t(\boldsymbol{X}|\boldsymbol{Y}_0;\ep)&=& \delta_{\boldsymbol{X},{\bf 0}}\left(1-\sum_{j=0}^t F_j({\bf
0}|\boldsymbol{Y}_0;\ep)\right) \nonumber\\
&+&\sum_{p=1}^{\infty}\sum_{m_1=1}^{\infty}\ldots\sum_{m_p=1}^{\infty}
\sum_{m_{p+1}=0}^{\infty}\delta_{m_1+\ldots+m_{p+1},t}\sum_{\nu_1}\ldots\sum_{\nu_p}\delta_{{\bf
e}_{{ \nu}_1}+\ldots+{\bf
e}_{{ \nu}_p},\boldsymbol{X}}  \nonumber\\
&\times& \left(1-\sum_{j=0}^{m_{p+1}}F_j({\bf
0}|-\ee_{\nu_p};\ep)\right)  F^*_{m_p}({\bf 0}|\ee_{\nu_p}|-\ee_{\nu_{p-1}};\ep)\ldots F^*_{m_2}({\bf 0}|\ee_{\nu_2}|-\ee_{\nu_1};\ep) F^*_{m_1}({\bf 0}|\ee_{\nu_1}|\boldsymbol{Y}_0;\ep).\nonumber\\
\label{eq:single_vac_prop_general}
\end{eqnarray}
The Fourier-Laplace transform of Eq. (\ref{eq:single_vac_prop_general}) writes
\begin{equation}
\label{singlevac}
\widetilde{\widehat{P}}^{(1)}(k|\ZZ;\xi,\epsilon)=\frac{1}{1-\xi}\left[1+\frac{1}{\mathcal{D}(k;\xi,\epsilon)}\sum_{b=\pm1} U_{b}(k;\xi) {F}^*(\zz|\ee_b|\ZZ;\xi,\epsilon)\right],
\end{equation}
where we defined:
\begin{eqnarray}
\mathcal{D}(k;\xi,\epsilon) & = & \mathrm{det}[\mathbf{1}-T(k;\xi,\epsilon)] \\
T(k;\xi,\epsilon) & = & \begin{pmatrix}
   \ex{\ii k}  \widehat{F}^*(\zz|\epu|\emu;\xi,\epsilon) &  \ex{\ii k}  \widehat{F}^*(\zz|\epu|\epu;\xi,\epsilon)   \\
   \ex{-\ii k}  \widehat{F}^*(\zz|\emu|\emu;\xi,\epsilon)    &   \ex{-\ii k}  \widehat{F}^*(\zz|\emu|\epu;\xi,\epsilon)
\end{pmatrix} \\
U_{b}(k;\xi,\epsilon) &=& \mathcal{D}(k;\xi,\epsilon)\sum_{{a}=\pm1}(1-\ex{-\ii a k})\left[  (\mathbf{1}-T(k;\xi,\epsilon))^{-1} \right]_{a,b}\ex{\ii b k}.
\end{eqnarray}
The TP being symmetric, we have the following relations
\begin{eqnarray}
\widehat{F}^*(\zz|\epu|\epu;\xi,\epsilon) & = & \widehat{F}^*(\zz|\emu|\emu;\xi,\epsilon) \\
\widehat{F}^*(\zz|\epu|\emu;\xi,\epsilon) & = & \widehat{F}^*(\zz|\emu|\epu;\xi,\epsilon),
\end{eqnarray}
so that $\mathcal{D}(k;\xi,\epsilon)$ and $U_{\pm 1}(k;\xi,\epsilon)$ reduce to:
\begin{eqnarray}
\mathcal{D}(k;\xi,\epsilon) & = & 1+\widehat{F}^*(\zz|\epu|\emu;\xi,\epsilon)^2-\widehat{F}^*(\zz|\epu|\epu;\xi,\epsilon)^2-2\widehat{F}^*(\zz|\epu|\emu;\xi,\epsilon)\cos k, \label{def_D} \\
U_{\pm1}(k;\xi,\epsilon) & = & (\ex{\pm\ii k}-1)[1-\ex{\mp \ii k}\widehat{F}^*(\zz|\epu|\emu;\xi,\epsilon)-\widehat{F}^*(\zz|\epu|\epu;\xi,\epsilon)].
\end{eqnarray}
We finally obtain the following expression of the single vacancy propagator in terms of the first-passage time densities $\widehat{F}^*$:
\begin{equation}
\label{P1vac_explicit}
\widetilde{\widehat{P}}^{(1)}(k|\ZZ;\xi,\epsilon)=\frac{1}{1-\xi}\left[1+\frac{\sum_{\nu=\pm1} F^*(\zz|\ee_\nu|\ZZ;\xi,\epsilon)(\ex{\nu\ii k}-1)\left(1-\ex{-\nu\ii k} \widehat{F}^*_{-1}-\widehat{F}^*_{1}\right)}{1-\left(\widehat{F}^*_1\right)^2+\left(\widehat{F}^*_{-1}\right)^2-2\widehat{F}^*_{-1}\cos k}\right],
\end{equation}
where we introduced the simplified notation
\begin{equation}
\widehat{F}^*_{\nu} \equiv F^*(\zz|\ee_1|\ee_{\nu};\xi,\epsilon).
\end{equation}

%In order to compute the single-vacancy propagator $\widehat{\widetilde{P}}^{(1)}(k|\ZZ;\xi,\epsilon)$, we simply need to compute the conditional propagators $F^*(\zz|\epu|\ee_{\pm1};\xi,\epsilon)$.

\subsection{Finite density of vacancies}

As for the study of the first regime, we now assume that the density of vacancies is equal to $\rho_0$, and we show that the Fourier-Laplace transform of the cumulant generating function in the limit $\rho_0\to 0$ is given by
\begin{equation}
\widehat{\Psi}(k;\xi,\epsilon) \underset{\rho_0\to0}{\sim} -\rho_0\sum_{a=\pm1} \left[ \frac{1}{1-\xi}  - \widehat{\widetilde{P}}^{(1)}(k|\ee_{-a};\xi,\epsilon)\ex{\ii a k} \right] \sum_{\ZZ\neq\zz} F^*(\zz|\ee_a|\ZZ;\xi,\epsilon).
\end{equation}
The TP being symmetric, the quantity $\sum_{\ZZ\neq\zz} F^*(\zz|\ee_a|\ZZ;\xi,\epsilon)$ is independent of $a$, 
\begin{equation}
\sum_{\ZZ\neq\zz} F^*(\zz|\ee_{-1}|\ZZ;\xi,\epsilon)=\sum_{\ZZ\neq\zz} F^*(\zz|\ee_1|\ZZ;\xi,\epsilon), 
\end{equation}
and we obtain
\begin{equation}
\widehat{\Psi}(k;\xi) \underset{\rho_0\to0}{\sim} -\rho_0 \left(  \sum_{\ZZ\neq\zz} F^*(\zz|\ee_1|\ZZ;\xi,\epsilon) \right)\sum_{a=\pm1} \left[ \frac{1}{1-\xi}  - \widehat{\widetilde{P}}^{(1)}(k|\ee_{-a};\xi,\epsilon)\ex{\ii a k} \right].
\end{equation}
Defining $\Sigma(\xi, \epsilon)=\sum_{\ZZ\neq\zz} F^*(\zz|\ee_1|\ZZ;\xi,\epsilon)$, we finally have
\begin{equation}
\label{Psi_final}
\widehat{\Psi}(k;\xi,\epsilon) \underset{\rho_0\to0}{\sim} -\rho_0 \Sigma(\xi,\epsilon)\sum_{a=\pm1} \left[ \frac{1}{1-\xi}  - \widehat{\widetilde{P}}^{(1)}(k|\ee_{-a};\xi,\epsilon)\ex{\ii a k} \right].
\end{equation}

The expression of the variance of the position of the TP is easily deduced from 
\begin{equation}
\widehat{\kappa}^{(2)}(\xi,\epsilon) =- \frac{\partial^2 \widehat{\Psi} (k;\xi,\epsilon)}{\partial k^2}.
\end{equation}
Using Eq. (\ref{P1vac_explicit}), we get
\begin{equation}
\widehat{\kappa}^{(2)}(\xi,\epsilon) = -2 \rho_0 \Sigma(\xi,\epsilon) \frac{\widehat{F}^*_{1}-\widehat{F}^*_{-1}-1}{\left(\widehat{F}^*_{1}-1+\widehat{F}^*_{-1}\right)   \left( \widehat{F}^*_{1}+1-\widehat{F}^*_{-1}\right)}.
\end{equation}
Consequently, the determination of the variance of the TP position only relies on the estimation on the following quantities:
\begin{itemize}
\item the sum $\Sigma(\xi,\epsilon)=\sum_{\ZZ\neq\zz} F^*(\zz|\ee_1|\ZZ;\xi,\epsilon)$,
\item the first-passage time densities $\widehat{F}^*_{\pm1} = \widehat{F}^*(\zz|\ee_1|\ee_{\pm1};\xi,\epsilon)$.
\end{itemize}
  In what follows, we evaluate the first-passage time densities $\widehat{F}^*_{\pm1}$ as well as the sum $\Sigma(\xi,\epsilon)$.

\subsubsection{Evaluation of $\widehat{F}^*(\zz|\ee_1|\ee_{\pm1};\xi,\epsilon)$}

In what follows, we write the conditional first-passage densities
$\widehat{F}^*(\zz|\ee_1|\ee_{\pm 1};\xi,\epsilon)$ in terms of the
propagators of the random walk of a vacancy with the rules specified
in Fig.~\ref{simplified_dynamics}. We first aim at computing
$\widehat{F}^*(\zz|\ee_1|\ee_{-1};\xi,\epsilon)$. We then assume that
the vacancy starts at site $\ee_{-1}$, and we write the following
partition over the visits of the vacancy to the origin of the lattice:
\begin{equation}
p(\ee_1\rightarrow \zz) P_{t-1}(\ee_1|\ee_{-1};\ep) = F_t^*(\zz|\ee_1|\ee_{-1};\ep) + \sum_{t'=0}^{t-1} p(\ee_1\rightarrow \zz) F_{t'}(\zz|\ee_{-1};\ep) P_{t-1-t'} (\ee_{-1}|\zz;\ep)
\end{equation}
where $p(\rr\rightarrow \rr')$ is the probability for a vacancy to jump from site $\rr$ to site $\rr'$ in a single step (here $p(\ee_1\rightarrow \zz)=1/4+\ep$). Multiplying the previous equation by $\xi^t$ and summing for $t$ going from $0$ to $\infty$, we obtain the following relation between the associated generating functions:
\begin{equation}
\xi \left( \frac{1}{4} + \ep\right) \widehat{P}(\ee_{1}|\ee_{-1};\xi,\ep) = \widehat{F}^*(\zz|\ee_{1}|\ee_{-1};\xi,\ep) + \xi \left( \frac{1}{4} + \ep\right)\widehat{ F}(\zz|\ee_{-1};\xi,\ep) \widehat{P}(\ee_1|\zz;\xi,\ep)
\end{equation}
The first-passage density $\widehat{ F}(\zz|\ee_{-1};\xi,\ep) $ is evaluated with the following renewal equation \cite{Hughes1995}:
\begin{equation}
\widehat{ F}(\zz|\ee_{-1};\xi,\ep) = \frac{\widehat{P}(\zz|\ee_{-1};\xi,\ep)}{\widehat{P}(\zz|\zz;\xi,\ep)}.
\end{equation}
We obtain the following expression of $\widehat{F}^*(\zz|\ee_{1}|\ee_{-1};\xi,\ep)$ in terms of the propagators $\widehat{P}$:
\begin{equation}
\widehat{F}^*(\zz|\ee_{1}|\ee_{-1};\xi,\ep) = \xi \left( \frac{1}{4} + \ep\right)  \left[    \widehat{P}(\ee_{1}|\ee_{-1};\xi,\ep) - \frac{\widehat{P}(\zz|\ee_{-1};\xi,\ep)\widehat{P}(\ee_1|\zz;\xi,\ep)   }{\widehat{P}(\zz|\zz;\xi,\ep)}    \right].
\end{equation}
The random walk performed by the vacancy being translationally invariant in the direction of the backbone, we will use the following relations:
\begin{eqnarray}
  \widehat{P}(\ee_{1}|\ee_{-1};\xi,\ep) & = &   \widehat{P}(2 \ee_{1}|\zz;\xi,\ep) \\
\widehat{P}(\zz|\ee_{-1};\xi,\ep) & = & \widehat{P}(\ee_1|\zz;\xi,\ep).
\end{eqnarray}
We define $\alpha(\rr;\xi,\ep)$ as
\begin{equation}
\alpha(\rr;\xi,\ep) \equiv \widehat{P}(\zz|\zz;\xi,\ep)- \widehat{P}(\rr|\zz;\xi,\ep).
\end{equation}
Then, the conditional first-passage densities are simply expressed
in terms of $\alpha(\ee_1;\xi,\ep)$, $\alpha(2 \ee_1;\xi,\ep)$ and
$\widehat{P}(\zz|\zz;\xi,\ep)$:
\begin{equation}
\label{exp_F11}
\widehat{F}^*(\zz|\ee_{1}|\ee_{-1};\xi,\ep) = \xi \left( \frac{1}{4} + \ep\right)  \left[    2 \alpha(\ee_1;\xi,\ep) - \alpha(\ee_1;\xi,\ep) -\frac{\alpha(\ee_1;\xi,\ep)^2 }{\widehat{P}(\zz|\zz;\xi,\ep)} \right].
\end{equation}

A similar calculation leads to the following expression of $\widehat{F}^*(\zz|\ee_{1}|\ee_{1};\xi,\ep)$ in terms of  $\alpha(\ee_1;\xi,\ep)$ and $\widehat{P}(\zz|\zz;\xi,\ep)$:
\begin{equation}
\label{exp_F1m1}
\widehat{F}^*(\zz|\ee_{1}|\ee_{1};\xi,\ep) = \xi \left( \frac{1}{4} + \ep\right)  \left[ 2 \alpha(\ee_1;\xi,\epsilon)-\frac{\alpha(\ee_1;\xi,\epsilon)^2}{\widehat{P}(\zz|\zz;\xi,\ep)}    \right].
\end{equation}

$\ $

In what follows, we compute the quantities $\alpha(\ee_1;\xi,\ep)$,  $\alpha(2 \ee_1;\xi,\ep)$ and $\widehat{P}(\zz|\zz;\xi,\ep)$ associated with the random walk of a vacancy with the evolution rules presented in Fig. \ref{simplified_dynamics}. We follow the method first introduced by Nieuwenhuizen et al. \cite{Nieuwenhuizen2004}.

For generality, we assume that if the vacancy is on the backbone (respectively not on the backbone), it has a probability $p_b^\parallel$ (resp. $p^\parallel$) to go left or right, and $p_b^\perp$ (resp. $p^\perp$) to go up or down. Let $P_{x,y}(t)$ be the probability to find the vacancy at site $(x,y)$ at time $t$, knowing that it started from the origin. The master equations of the problem are the following:
\begin{itemize}
  \item if $y\neq0,\pm1$:
  \begin{equation}
P_{x,y}(t+1)=p^\parallel[P_{x+1,y}(t)+P_{x-1,y}(t)]+p^\perp[P_{x,y+1}(t)+P_{x,y-1}(t)]
\end{equation}
  \item if $y=0$:
    \begin{equation}
P_{x,0}(t+1)=p^\perp[P_{x,1}(t)+P_{x,-1}(t)]+p_b^\parallel[P_{x+1,0}(t)+P_{x-1,0}(t)]
\end{equation}
  \item if $y=\pm1$:
      \begin{equation}
P_{x,\pm1}(t+1)=p^\perp P_{x,\pm2}(t)+p_b^\perp P_{x,0}(t)+p^\parallel[P_{x+1,\pm1}(t)+P_{x-1,\pm1}(t)].
\end{equation}
\end{itemize}
We introduce the following generating functions and Laplace transforms:
\begin{eqnarray}
P_b(k_x;\xi) & \equiv & \sum_{t=0}^\infty  \sum_{x=-\infty}^\infty \ex{\ii k_x x} P_{x,0}(t)\xi^t, \label{def_Pb}\\
P(k_x,k_y;\xi) & \equiv &  \sum_{t=0}^\infty  \sum_{x,y=-\infty}^\infty \ex{\ii k_x x} \ex{\ii k_y y}  P_{x,y}(t)\xi^t .
\end{eqnarray}
Using the master equations, it is straightforward to show that $P$ and $P_b$ are related by
\begin{equation}
P(k_x,k_y;\xi)=\frac{1+2\xi[(p_b^\perp-p^\perp)\cos k_y +(p_b^\parallel-p^\parallel)\cos k_x]P_b(k_x;\xi)}{1-2\xi[p^\perp\cos k_y +p^\parallel \cos k_x]}.
\end{equation}
In order to get an equation for $P_b(k_x;\xi)$, we integrate each side of this equation over $k_y$ and use the simple relation between $P$ and $P_b$:
\begin{equation}
P_b(k_x;\xi)=\int_0^{2\pi} \frac{\dd k_y}{2\pi} P(k_x,k_y;\xi).
\end{equation}
We also use the following useful integrals:
\begin{eqnarray}
\int_0^{2\pi} \frac{\dd k}{2\pi} \frac{1}{\cosh \mu-\cos k}  & = &\frac{1}{\sinh \mu},  \\
\int_0^{2\pi} \frac{\dd k}{2\pi} \frac{\cos k}{\cosh \mu-\cos k} & = & \frac{\ex{-\mu}}{\sinh \mu}, 
\end{eqnarray}
to finally obtain
\begin{equation}
\label{Pb_1}
P_b(k_x;\xi)=\frac{1}{2\xi p^\perp \sinh \mu-2\xi[(p_b^\perp-p^\perp)\ex{-\mu}+(p_b^\parallel-p^\parallel)\cos k_x]},
\end{equation}
where $\mu$ is such that
\begin{equation}
\cosh \mu = \frac{1-2\xi p^\parallel\cos k_x}{2\xi p^\perp}.
\end{equation}
In what follows, we will use the explicit expressions of the transition probabilities:
\begin{eqnarray}
p^\parallel & = & \epsilon, \\
p^\perp & = & \frac{1}{2}-\epsilon, \\ 
p_b^\parallel & = & \frac{1}{4}+\epsilon, \\ 
p_b^\perp & = & \frac{1}{4}-\epsilon \label{Pb_end}.
\end{eqnarray}

\paragraph{Computation of $\alpha(2\ee_1;\xi,\epsilon)$}

 We focus on the computation of $\alpha(2\ee_1;\xi,\epsilon)$, defined by:
\begin{eqnarray}
\alpha(2\ee_1;\xi,\epsilon)&=&\widehat{P}(\zz|\zz;\xi,\epsilon)-\widehat{P}(2\ee_1|\zz;\xi,\epsilon) \\
&=&\int_0^{2\pi}\frac{\dd k_x}{2\pi} (1-\ex{-2\ii k_x})P_b(k_x;\xi).
\end{eqnarray}
In the limit where we first take $\xi=1$, one has:
\begin{equation}
\label{exp_Pb}
P_b(k_x;\xi=1)=\frac{2(1-2\epsilon)}{1-\cos k_x +2\sqrt{\epsilon}(1-4\epsilon)\sqrt{(1-\cos k_x)(1-\epsilon-\epsilon\cos k_x)}}
\end{equation}
and consequently
\begin{equation}
\alpha(2\ee_1;\xi=1,\epsilon) = \int_0^{2\pi} \frac{\dd k_x}{2\pi} \frac{2(1-2\epsilon)(1-\ex{-2\ii k_x})}{1-\cos k_x +2\sqrt{\epsilon}(1-4\epsilon)\sqrt{(1-\cos k_x)(1-\epsilon-\epsilon\cos k_x)}}.
\end{equation}
For symmetry reasons, the imaginary part of the integral is zero, and 
\begin{equation}
\alpha(2\ee_1;\xi=1,\epsilon) = \frac{4}{\pi}\int_0^{\pi} \dd k_x \frac{(1-2\epsilon)\sin^2 k_x}{1-\cos k_x +2\sqrt{\epsilon}(1-4\epsilon)\sqrt{(1-\cos k_x)(1-\epsilon-\epsilon\cos k_x)}}.
\end{equation}
With the change of variable $u=\cos k_x$, one gets
\begin{eqnarray}
\alpha(2\ee_1;\xi=1,\epsilon) &=& \frac{4}{\pi}(1-2\epsilon)\int_{-1}^{1} \dd u \frac{\sqrt{1-u^2}}{1-u +2\sqrt{\epsilon}(1-4\epsilon)\sqrt{(1-u)(1-\epsilon-\epsilon u)}}\\
   &=& \frac{4}{\pi}(1-2\epsilon)\int_{-1}^{1} \dd u \frac{\sqrt{1+u}}{\sqrt{1-u} +2\sqrt{\epsilon}(1-4\epsilon)\sqrt{1-\epsilon-\epsilon u}}.
\end{eqnarray}

We aim to compute the expansion of $\alpha(2\ee_1|\zz;\xi=1,\epsilon)$ in powers of $\epsilon$. The integral in the previous equation have an explicit expression:
\begin{equation}
\begin{split}
\alpha(2\ee_1;\xi=1,\epsilon)
&=\frac{4(1-2\epsilon)}{\pi(2 \epsilon-1) \left(32 \epsilon^3+2 \epsilon+1\right)^2}[256 \sqrt{2} \sqrt{(1-2 \epsilon) \epsilon^9}-64 \sqrt{2} \sqrt{(1-2 \epsilon) \epsilon^7}
16   \sqrt{2} \sqrt{(1-2 \epsilon) \epsilon^5}+32 \pi  \epsilon^3\\
&\quad+4 \sqrt{2} \sqrt{(1-2 \epsilon)
   \epsilon^3}+2 \ii \sqrt{2} \pi  \sqrt{16 \epsilon^3+\epsilon}-32 \pi  \epsilon^2-2 (1-4 \epsilon)^2
   \left(8 \epsilon^2+2 \epsilon+1\right) \arctan \left(\sqrt{2} \sqrt{\frac{\epsilon}{1-2
   \epsilon}}\right)\\
   &\quad -8 \ii \sqrt{2} \pi  \sqrt{16 \epsilon^5+\epsilon^3}+4 \sqrt{2} \left(\sqrt{16
   \epsilon^3+\epsilon}-4 \sqrt{16 \epsilon^5+\epsilon^3}\right) \mathrm{arctanh} \left(\frac{1}{\sqrt{-32
   \epsilon^3+16 \epsilon^2-2 \epsilon+1}}\right)\\
 &  \quad+2 \pi  \epsilon-2 \sqrt{2} \sqrt{(1-2 \epsilon)
   \epsilon}-\pi ].
\end{split}
\end{equation}
Its expansion for $\ep\to0$ leads to:
\begin{equation}
\label{alpha2_exp}
\alpha(2\ee_1;\xi=1,\epsilon)\underset{\epsilon\to 0}{=} 4+\frac{8\sqrt{2}}{\pi}\sqrt{\epsilon}\ln \epsilon+\cdots
\end{equation}

%A  straightforward way to estimate the behavior of the integral in the limit where $\epsilon\to0$ is to use its approximate expression:
%\begin{eqnarray}
%\alpha(2\ee_1;\xi=1,\epsilon) &\simeq& \frac{4}{\pi}\int_{-1}^{1} \dd u \frac{\sqrt{1+u}}{1-u +2\sqrt{\epsilon}}\\
%&=&4-\frac{16}{\pi}\sqrt{2\epsilon}\sqrt{1-2\epsilon}\,\mathrm{arccosh}\frac{1}{\sqrt{2\epsilon}}+\frac{16\sqrt{2}}{\pi}\sqrt{\epsilon}-16\epsilon \\
%&\underset{\epsilon\to0}{=}&4+\frac{8\sqrt{2}}{\pi}\sqrt{\epsilon}\ln \epsilon+\cdots
%\end{eqnarray}

\paragraph{Computation of $\alpha(\ee_1|\zz;\xi=1,\epsilon)$} 

Similarly, we study $\alpha(\ee_1|\zz;\xi=1,\epsilon)$, defined by
\begin{eqnarray}
\alpha(\ee_1;\xi,\epsilon)&=&\widehat{P}(\zz|\zz;\xi,\epsilon)-\widehat{P}(\ee_1|\zz;\xi,\epsilon) \\
&=&\int_0^{2\pi}\frac{\dd k_x}{2\pi} (1-\ex{-\ii k_x})P_b(k_x;\xi).
\end{eqnarray}
Using again Eq. (\ref{exp_Pb}), and with the change of variable $u=\cos k_x$, we obtain
\begin{equation}
\alpha(\ee_1;\xi,\epsilon)=  \frac{2}{\pi} (1-2\ep) \int_{-1}^1 \dd u \frac{1}{\sqrt{1-u^2} + 2\sqrt{\ep}(1-4\ep)\sqrt{(1+u)(1-\ep-\ep u)}}  
\end{equation}
The explicit expression of this integral is
\begin{equation}
\begin{split}
\alpha(\ee_1;\xi=1,\epsilon)
&=\frac{\sqrt{2}}{\pi(1+4\ep)(1-2\ep+8\ep^2)\sqrt{\ep}\sqrt{1+16\ep^2}}[\pi\sqrt{1+16\ep^2}\sqrt{2\ep}(2\ep^2-\ep+1) + 4\ii\pi\ep(4\ep-1)\\
&\quad+\mathrm{Arcsin}(4\ep-1)\sqrt{1+16\ep^2}2\sqrt{2}\ep^{3/2}(4\ep-1) +2\ep (1-4\ep)\ln\ep+4\ep(1-4\ep)\ln(1-4\ep) \\
&\quad+2\ep(4\ep-1) \ln(-16{\epsilon}^{3}+\sqrt {1-2\epsilon}\sqrt {16{\epsilon}^{2}+1}+8{\epsilon}^{2}-\epsilon+1)].
\end{split}
\end{equation}
Its expansion for $\ep\to0$ leads to:
\begin{equation}
\label{alpha1_exp}
\alpha(\ee_1;\xi=1,\epsilon)\underset{\epsilon\to 0}{=} 2+\frac{2\sqrt{2}}{\pi}\sqrt{\epsilon}\ln \epsilon+\cdots
\end{equation}

\paragraph{Computation of $\widehat{P}(\zz|\zz;\xi,\epsilon)$}

Using the definition of $P_b$ from Eq.~(\ref{def_Pb}), the propagator $\widehat{P}(\zz|\zz;\xi,\epsilon)$ can be written
\begin{equation}
\widehat{P}(\zz|\zz;\xi,\epsilon) = \int_0^{2\pi} \frac{\dd k_x}{2\pi} P_b(k_x;\xi),
\end{equation}
where $P_b(k_x;\xi)$ is defined by Eqs. (\ref{Pb_1}) to (\ref{Pb_end}). We then express $\widehat{P}(\zz|\zz;\xi,\epsilon)$ as the following integral
\begin{equation}
\widehat{P}(\zz|\zz;\xi,\epsilon) =  \int_0^{2\pi} \frac{\dd k_x}{2\pi} \frac{2(1-2\ep)}{1-\xi \cos k_x +(1-4\ep)\sqrt{(1-2\xi\ep\cos k_x)^2-\xi^2(1-2\ep)^2}}.
\end{equation}
With the change of variable $u=\cos k_x$, this integral is rewritten as
\begin{equation}
\widehat{P}(\zz|\zz;\xi,\epsilon) = \frac{2}{\pi} \int_0^{\pi} \frac{\dd u}{\sqrt{1-u^2}} \frac{1-2\ep}{1-\xi u +(1-4\ep)\sqrt{(1-2\xi\ep u)^2-\xi^2(1-2\ep)^2}}.
\end{equation}
In the limit where $\xi \to 1$ and $\ep \to 0$ this quantity diverges. Studying the integrand, we show that this divergence is located near $u=1$. Introducing a new integration variable $v=1-u$, we obtain that, for $\xi\to1$,  the integral is estimated at leading order by
\begin{equation}
\widehat{P}(\zz|\zz;\xi,\epsilon) \underset{\xi \to 1}{\sim} \frac{1-2\ep}{\pi \ep (1-4\ep)} \int_0^{2} \frac{\dd v}{\sqrt{v}\sqrt{2-v}\sqrt{v+\frac{1-\xi}{2\ep}}  \sqrt{v+\frac{1}{\ep}-2}}.
\end{equation}
The integral over $v$ is conveniently expressed in terms of the complete elliptic integral of the first kind $\mathrm{K}(x)$:
\begin{equation}
\widehat{P}(\zz|\zz;\xi,\epsilon) \underset{\xi \to 1}{\sim} \frac{1-2\ep}{\pi \ep (1-4\ep)} \frac{2\sqrt{2}\ep}{\sqrt{1-\xi}} \mathrm{K}\left(\frac{2\ep(4\ep-1-\xi)}{1-\xi} \right).
\end{equation}
We use the following useful expansion of $\mathrm{K}(x)$:
\begin{equation}
\mathrm{K}\left(-\frac{1}{x} \right) \underset{x\to 0^+}{\sim} \frac{\sqrt{x}}{2} \ln \frac{1}{x},
\end{equation}
to finally obtain the expansion of $\widehat{P}(\zz|\zz;\xi,\epsilon)$ at leading order in $(1-\xi)$:
\begin{equation}
\label{P00_exp}
\widehat{P}(\zz|\zz;\xi,\epsilon) \underset{\xi \to 1}{\sim} \frac{\sqrt{1-2\ep}}{\pi\sqrt{2 \ep} (1-4\ep)} \ln \frac{1}{1-\xi}.
\end{equation}

Finally, recalling the expressions of $\widehat{F}^*(\zz|\ee_1|\ee_{\pm1};\xi,\ep)$ in terms of the functions $\alpha$ and of the propagator $\widehat{P}(\zz|\zz;\xi,\epsilon)$ (Eqs. (\ref{exp_F11}) and (\ref{exp_F1m1})), we use the results from Eqs. (\ref{alpha1_exp}), (\ref{alpha2_exp}) and (\ref{P00_exp}), we obtain the following expansions
\begin{equation}
\widehat{F}^*(\zz|\ee_1|\ee_{\pm1};\xi,\ep)  \underset{\xi \to 1}{=}  A_{\pm1}(\ep)+C(\ep) \frac{1}{\ln \frac{1}{1-\xi}} + \dots,
\end{equation}
where $A_{\pm1}(\ep)$ and $C(\ep)$ have the following expansions when $\ep\to0$:
\begin{eqnarray}
A_{1}(\ep) & \underset{\ep\to0}{\sim} &1 \\
A_{-1}(\ep) & \underset{\ep\to0}{\sim} & \frac{\sqrt{2}}{\pi} \sqrt{\ep} \ln \frac{1}{\ep} \\ 
C(\ep) & \underset{\ep\to0}{\sim} & -\pi \sqrt{\ep}.
\end{eqnarray}

Recalling that the additional jump probability $\epsilon=D(\rho_0)$ is a vanishing function of $\rho$ when $rho_0\to0$, the conditional first-passage densities can be rewritten in terms of $\rho_0$ as follows:
\begin{equation}
\label{calcul_FPTD}
\widehat{F}^*(\zz|\ee_1|\ee_{\pm1};\xi,\rho_0)  \underset{\xi \to 1}{=}  A_{\pm1}(\rho_0)+C(\rho_0) \frac{1}{\ln \frac{1}{1-\xi}} + \dots,
\end{equation}
where $A_{\pm1}(\rho_0)$ and $C(\rho_0)$ have the following expansions when $\rho_0\to0$:
\begin{eqnarray}
A_{1}(\rho_0) & \underset{\ep\to0}{\sim} &1 \\
A_{-1}(\rho_0) & \underset{\ep\to0}{\propto} & \frac{\sqrt{2}}{\pi} \sqrt{D(\rho_0)} \ln \frac{1}{D(\rho_0)} \\ 
C(\rho_0) & \underset{\ep\to0}{\propto} & -\pi \sqrt{D(\rho_0)}.
\end{eqnarray}

\subsubsection{Evaluation of $\Sigma(\xi,\epsilon)$}

The evaluation of the sum $\Sigma(\xi,\epsilon)$ is conveniently done
by adopting a continuous-space description of the system. Following
the approach proposed by Arkhincheev and
Baskin~\cite{Arkhincheev1991,Arkhincheev2002}, we define by
$G(x,y|x_0,y_0;t)$ the propagator associated with the random walk of a
vacancy starting from site $(x_0,y_0)$ and arriving at $(x,y)$. The
diffusion equation verified by $G$ is then
\begin{equation}
\label{eq:cont_diff_eq}
\begin{dcases}
   \frac{\partial G}{\partial t} = \left[  D_1 \delta (y) + D(\rho_0) \right] \frac{\partial^2 G}{\partial x^2} + D_2 \frac{\partial^2G}{\partial y^2},    \\
  G(x,y|x_0,y_0;t=0)=\delta(x-x_0)\delta(y-y_0),
\end{dcases}
\end{equation}
where $D_1$ (resp. $D_2$) is the diffusion coefficient of the vacancy in the $x$ (resp. $y$) direction, and $D(\rho_0)$ is the additional jump probability experienced by the vacancy due to the effective diffusion of the TP.  This coefficient is not explicitly known, and is only assumed to vanish in the limit $\rho_0 \to 0$. 
%It will be determined self-consistently at the end of the calculation.

We now solve the diffusion equation (\ref{eq:cont_diff_eq}) in order to obtain an expression for $G(x,y|x_0,y_0;t)$. We introduce the continuous Laplace transform, defined for any time-dependent function by 
\begin{equation}
\widehat{f}(s)=\int_0^\infty \dd t \, \ex{-st} f(t),
\end{equation}
and the Fourier transform along the $x$-direction, defined for any $x$-dependent function:
\begin{equation}
\widetilde{g}(k)=\int_{-\infty}^\infty \dd x \, \ex{-\ii k x} g(x).
\end{equation}
The Fourier-Laplace transform of (\ref{eq:cont_diff_eq}) is
\begin{equation}
\label{eq:TF_TL_cont}
\left[s + D(\rho_0) k^2-D_2 \frac{\partial^2}{\partial y^2}   \right] \widehat{\widetilde{G}} (k,y|x_0,y_0;s)=\ex{-\ii k x_0} \delta(y-y_0).
\end{equation}
We first assume that $y_0>0$. On each of the  three domains $]-\infty,0[$, $]0,y_0[$ and $]y_0,\infty[$, Eq. (\ref{eq:TF_TL_cont}) becomes
\begin{equation}
\label{eq:TF_TL_cont_2}
\left[s + D(\rho_0) k^2-D_2 \frac{\partial^2}{\partial y^2}   \right] \widehat{\widetilde{G}} (k,y|x_0,y_0;s)=0,
\end{equation}
whose general solution is
\begin{equation}
\widehat{\widetilde{G}} (k,y|x_0,y_0;s) = A\exp\left({-\sqrt{\frac{s+D(\rho_0)k^2}{D_2}}}\right)+B\exp\left({\sqrt{\frac{s+D(\rho_0)k^2}{D_2}}}\right),
\end{equation}
where $A$ and $B$ are two constants to be determined. The physically relevant solutions on the three domains take the form
\begin{equation}
\widehat{\widetilde{G}} (k,y|x_0,y_0;s) =
 \begin{cases}
  B_1 \exp\left({\sqrt{\frac{s+D(\rho_0) k^2}{D_2}} y}  \right)  & \text{if $y<0$}, \\
   A_2 \exp\left(-{\sqrt{\frac{s+D(\rho_0) k^2}{D_2}} y}  \right)  +B_2 \exp\left({\sqrt{\frac{s+D(\rho_0) k^2}{D_2}} y}  \right) & \text{if $0<y<y_0$}, \\
 A_3 \exp\left(-{\sqrt{\frac{s+D(\rho_0) k^2}{D_2}} y}  \right)   & \text{if $y_0<y$}.
\end{cases}
\end{equation}
There are four constants ($B_1$, $A_2$, $B_2$ and $A_3$) to be determined. The continuity of $G$ as a function of $y$ yields the following relation:
\begin{equation}
\begin{cases}
B_1=A_2+B_2,\\
A_3 = A_2+B_2\exp\left(2{\sqrt{\frac{s+D(\rho_0) k^2}{D_2}} y}  \right), 
\end{cases}
\end{equation}
so that we write
\begin{equation}
\widehat{\widetilde{G}} (k,y|x_0,y_0;s) =
 \begin{cases}
  (A_2+B_2) \exp\left({\sqrt{\frac{s+D(\rho_0) k^2}{D_2}} y}  \right)  & \text{if $y<0$}, \\
   A_2 \exp\left(-{\sqrt{\frac{s+D(\rho_0) k^2}{D_2}} y}  \right)  +B_2 \exp\left({\sqrt{\frac{s+D(\rho_0) k^2}{D_2}} y}  \right) & \text{if $0<y<y_0$}, \\
 A_2 \exp\left(-{\sqrt{\frac{s+D(\rho_0) k^2}{D_2}} y}  \right) +   B_2  \exp\left(\sqrt{\frac{s+D(\rho_0) k^2}{D_2}} (2y_0-y)  \right)& \text{if $y_0<y$}.
\end{cases}
\end{equation}
The constants $A_2$ and $B_2$ are obtained by integrating Eq.~(\ref{eq:TF_TL_cont_2}) respectively over the intervals $[-\eta,\eta]$ and $[y_0-\eta,y_0+\eta]$ and by taking the limits $\eta\to0$. On the intervals $]-\infty,0[$ and $]0,y_0[$, the following relation holds
\begin{equation}
\widehat{\widetilde{G}} (k,y|x_0,y_0;s) = A_2 \exp\left(-{\sqrt{\frac{s+D(\rho_0) k^2}{D_2}} |y|}  \right) +B_2 \exp\left({\sqrt{\frac{s+D(\rho_0) k^2}{D_2}} y}  \right).
\end{equation}
We integrate Eq.~(\ref{eq:TF_TL_cont_2}) over $[-\eta,\eta]$ and using the following general relation (which holds for any $\alpha>0$)
\begin{equation}
\label{general_int}
\int_{-\eta}^\eta \dd y\,  \frac{\partial^2}{\partial y^2 }\left[  \ex{-\alpha|y|} \right]  =  \int_{-\eta}^\eta \dd y\, \alpha[\alpha-2\delta(y)] \ex{-\alpha |y|} \underset{\eta\to0}{\rightarrow}  -2 \alpha,
\end{equation}
we obtain the relation
\begin{equation}
\label{rel1_A2B2}
D_1 k^2 (A_2+B_2)+2A_2 \sqrt{D_2[s+D(\rho_0) k^2]} = 0.
\end{equation}
On the intervals $]0,y_0[$ and $[y_0,\infty[$, the following expression of $\widehat{\widetilde{G}}$ holds:
\begin{equation}
\widehat{\widetilde{G}} (k,y|x_0,y_0;s) = A_2 \exp\left(-{\sqrt{\frac{s+D(\rho_0) k^2}{D_2}} y}  \right) +B_2 \exp\left( -\sqrt{\frac{s+D(\rho_0)k^2}{D_2}} y_0 \right)    \exp\left( -\sqrt{\frac{s+D(\rho_0)k^2}{D_2}} |y- y_0| \right).
\end{equation}
We integrate Eq. (\ref{eq:TF_TL_cont_2}) over $[y_0- \eta,y_0 + \eta]$, and using again Eq. (\ref{general_int}), we obtain the expression of $B_2$:
\begin{equation}
B_2 = \frac{1}{2\sqrt{D_2[s+D(\rho_0) k^2]}} \exp \left( -\ii k x_0  -\sqrt{\frac{s+D(\rho_0)k^2}{D_2}} y_0 \right).
\end{equation}
The relation between $A_2$ and $B_2$ yields
\begin{equation}
A_2=- \frac{D_1 k^2}{D_1 k^2 +2\sqrt{D_2[s+D(\rho_0) k^2]}}   \frac{1}{2\sqrt{D_2[s+D(\rho_0) k^2]}} \exp \left( -\ii k x_0  -\sqrt{\frac{s+D(\rho_0)k^2}{D_2}} y_0 \right).
\end{equation}
In what follows, we will only consider the propagator of a random walk starting from a generic point $(x_0,y_0)$ and arriving at the origin $(0,0)$, so that we will take $y=0$. Generalizing the calculation to the case of a starting point located in the domain $y_0<0$, we finally obtain the general expression
\begin{equation}
\label{G_Lap_Four}
\widehat{\widetilde{G}} (k,0|x_0,y_0;s) =\frac{\exp\left( -\ii k x_0 - \sqrt{\frac{s+D(\rho_0) k^2}{D_2}} | y_0|\right)}{D_1 k^2+2\sqrt{D_2[s+ D(\rho_0) k^2]}}.
\end{equation}

We now study the asymptotic behavior of the sum $\Sigma(\xi,\epsilon) = \sum_{\ZZ\neq\zz} \widehat{F}^*(\zz|\ee_1|\ZZ;\xi,\ep)$ in the limit where $\xi\to 1$ (long-time limit). In the continuous-space and time description, we replace the sum by the following integral
\begin{equation}
\Sigma(s,\rho_0) = \int \dd x_0 \int  \dd y_0 \,  \widehat{F}(0,0|x_0,y_0;s).
\end{equation}
Writing the following renewal equation
\begin{equation}
\widehat{F}(0,0|x_0,y_0;s) = \frac{\widehat{G}(0,0|x_0,y_0;s)}{\widehat{G}(0,0|0,0;s)},
\end{equation}
we get
\begin{equation}
\label{Sigma_cont}
\Sigma(s,\rho_0) = \frac{1 }{\widehat{G}(0,0|0,0;s)}\int \dd x_0 \int \dd y_0 \,  \widehat{G}(0,0|x_0,y_0;s).
\end{equation}
Taking the inverse Fourier transform of  Eq. (\ref{G_Lap_Four}), we write
\begin{equation}
\label{G_Lap}
\widehat{G}(x,0|x_0,y_0;s) = \int_{-\infty}^\infty   \frac{\dd k}{2\pi }   \exp(\ii k x) \frac{\exp\left( -\ii k x_0 - \sqrt{\frac{s+D(\rho_0) k^2}{D_2}} | y_0|\right)}{D_1 k^2+2\sqrt{D_2[s+ D(\rho_0) k^2]}}.
\end{equation}
We now evaluate separately $\widehat{G}(0,0|0,0;s)$ and $\int \dd x_0 \,  \dd y_0 \,  \widehat{G}(0,0|x_0,y_0;s)$:
\begin{itemize}
  \item using Eq. (\ref{G_Lap}), we obtain the expression of $\widehat{G}(0,0|0,0;s)$:
  \begin{eqnarray}
\widehat{G}(0,0|0,0;s) & = & \int_{-\infty}^\infty   \frac{\dd k}{2\pi }  \frac{1}{D_1 k^2+2\sqrt{D_2[s+ D(\rho_0) k^2]}}, \\
 & = & \frac{1}{\pi} \int_0^\infty \dd k \frac{1}{D_1 k^2+2\sqrt{D_2[s+ D(\rho_0) k^2]}}.
\end{eqnarray}
In the long-time limit ($s\to0$), the integral is dominated by the small values of the variable $k$. Introducing a cutoff value $A$, we write
\begin{equation}
\widehat{G}(0,0|0,0;s)   \sim  \frac{1}{\pi} \int_0^A \dd k \frac{1}{2\sqrt{D_2[s+ D(\rho_0) k^2]}}. \\
\end{equation}
Using the following integral (for $a>0$),
\begin{equation}
\int_0^A \dd x\, \frac{1}{\sqrt{a^2+x^2}} = \ln \left( \frac{A}{a}+ \sqrt{\frac{A}{a}+1} \right),
\end{equation}
we obtain the leading order behavior of $\widehat{G}(0,0|0,0;s) $ when $s \to 0$:
\begin{equation}
\label{eval_fct_Green}
\widehat{G}(0,0|0,0;s)  \underset{s \to 0}{\sim} \frac{1}{4\pi \sqrt{D_2}} \frac{1}{\sqrt{D(\rho_0)}} \ln \frac{1}{s}.
\end{equation}
  \item the integral $\int \dd x_0 \int  \dd y_0 \,  \widehat{G}(0,0|x_0,y_0;s)$, using Eq. (\ref{G_Lap}), yields:
  \begin{eqnarray}
\int \dd x_0 \int  \dd y_0 \,  \widehat{G}(0,0|x_0,y_0;s) & = & \int \dd x_0 \int  \dd y_0 \int_{-\infty}^\infty \frac{\dd k}{2\pi}  \frac{\exp\left( -\ii k x_0 - \sqrt{\frac{s+D(\rho_0) k^2}{D_2}} | y_0|\right)}{D_1 k^2+2\sqrt{D_2[s+ D(\rho_0) k^2]}}   \\
 & = & \int_{-\infty}^\infty \frac{\dd k }{\pi } \int \dd x_0 \frac{\exp(-\ii k x_0)}{D_1 k^2+2\sqrt{D_2[s+ D(\rho_0) k^2]}} \sqrt{\frac{D_2}{s+ D(\rho_0) k^2}} \\
 &=&  \int_{-\infty}^\infty \dd k   \frac{2\delta(k)}{D_1 k^2+2\sqrt{D_2[s+ D(\rho_0) k^2]}} \sqrt{\frac{D_2}{s+ D(\rho_0) k^2}} \\
 & =& \frac{1}{s} \label{eval_int_pts_depart}
\end{eqnarray}
\end{itemize}
Finally, using the estimates from Eqs. (\ref{eval_fct_Green}) and (\ref{eval_int_pts_depart}) in Eq. (\ref{Sigma_cont}), we get
\begin{equation}
\Sigma(s,\rho_0) \underset{s \to 0}{\sim} \frac{4\pi \sqrt{D_2} \sqrt{D(\rho_0)}}{s \ln (1/s)}.
\end{equation}
In the long-time limit, we get the equivalent discrete time description in terms of the variable $\xi$:
\begin{equation}
\label{calcul_sum}
\Sigma(\xi,\rho_0) \underset{\xi \to 1}{\sim} \frac{4\pi \sqrt{D_2} \sqrt{D(\rho_0)}}{(1-\xi) \ln \frac{1}{1-\xi}}.
\end{equation}

\subsubsection{Expression of the variance}

\begin{figure}
\begin{center}
\includegraphics[width=8.5cm,clip=true]{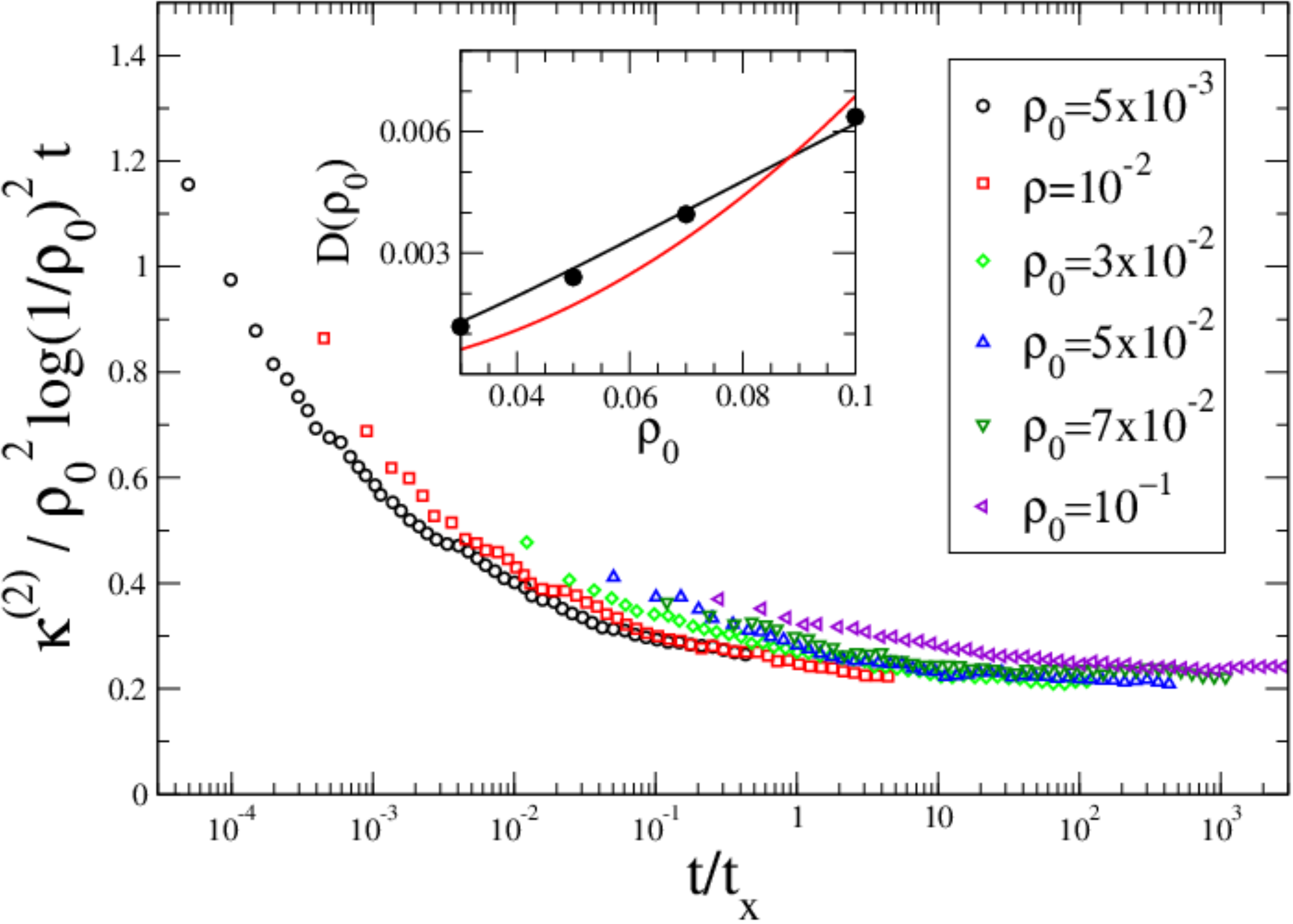}
\caption{(Color online) Variance of the TP when its motion is
  constrained on the backbone, rescaled by the time $t$ and $\rho_0^2\log(1/\rho_0)^2$, according to Eq.~(11) of the main text, for several values
  of $\rho_0$. Inset: the black dots represent the diffusion coefficient $D(\rho_0)=\lim_{t\to \infty}\kappa^{(2)}(t)/2t$ measured in simulations and the black line is the best fit
obtained with the function $f(x)=ax^2\log(1/x)^2$ (a=0.2336). For comparison we also show the best fit obtained with the function $g(x)=bx^2$, with $b=1.375$ (red line).}
\label{scaling_log}
\end{center}
\end{figure}

We recall the expression of the generating function associated with the variance of the TP position:
\begin{equation}
\widehat{\kappa}^{(2)}(\xi,\rho_0) = -2 \rho_0 \Sigma(\xi,\rho_0) \frac{\widehat{F}^*_{1}-\widehat{F}^*_{-1}-1}{\left(\widehat{F}^*_{1}-1+\widehat{F}^*_{-1}\right)   \left( \widehat{F}^*_{1}+1-\widehat{F}^*_{-1}\right)}.
\end{equation}
Using the results from Eqs. (\ref{calcul_sum}) and (\ref{calcul_FPTD}), we obtain the  expansion of the generating function associated with the variance up to a numerical prefactor, where we first take the long-time limit ($\xi\to1$) and ultimately the high-density limit ($\rho_0\to0$):
\begin{equation}
\widehat{\kappa}^{(2)}(\xi,\rho_0) \propto \frac{\rho_0}{(1-\xi)^2} \sqrt{D(\rho_0)} \ln \frac{1}{D(\rho_0)}.
\end{equation}
Using a Tauberian theorem, we retrieve the time-dependence of the cumulant and the expression presented in the main text:
\begin{equation}
%\label{variance_diff}
\lim_{t \to \infty} \frac{\kappa^{(2)}(t)}{t}   \underset{\rho_0 \to 0}{\propto} \rho_0  \sqrt{D(\rho_0)} \ln \frac{1}{D(\rho_0)}.
\end{equation}
From this expression we obtain Eq.~(10) of the main text, which turns to be in good agreement with numerical simulations,
as shown in Fig.~\ref{scaling_log}.

\section{Numerical simulations}

For the case where the TP is constrained on the backbone, the data
reported are obtained via Monte Carlo numerical simulations of vacancy
dynamics. We considered a lattice of size $L_x\times L_y=1000^2$ with
periodic boundary conditions and a number of vacancies $M=\rho_0 L_x
L_y$. At each Monte Carlo step, the position of all vacancies is
updated according to the evolution rules described in
Section~\ref{evol_rules}. If a vacancy exchanges its position with the
TP, the TP position is also updated and the displacement is
measured. Reported data are averaged over some thousands of
realizations.

In the case where the TP is allowed to visit the teeth of the lattice,
for values of $\rho_0\le 10^{-3}$ the reported data are obtained via
Monte Carlo simulations of vacancy dynamics, with the same lattice
size as above. For larger values of $\rho_0$ we performed Monte Carlo
simulations of particle dynamics. At each Monte Carlo step, all
particles (included the TP) attempt a move on a neighbor site with the
probabilities given in the main text, and the move is accepted if the
target site is empty. In this case, the lattice size is $L_x\times L_y=200^2$
with periodic boundary conditions and the number of particles is
$N=(1-\rho_0)L_xL_y$.

% \begin{thebibliography}{10}

% \bibitem{Woess2000}
% Wolfgang Woess.
% \newblock {\em {Random Walks on Infinite Graphs and Groups}}.
% \newblock Cambridge University Press, 2000.

% \bibitem{Hughes1995}
% Barry~D. Hughes.
% \newblock {\em {Random Walks and Random Environments: Random walks, Volume 1}}.
% \newblock Oxford University, New York, 1995.

% \bibitem{Nieuwenhuizen2004}
% Theo Nieuwenhuizen, Stefan Klumpp, and Reinhard Lipowsky.
% \newblock {Random walks of molecular motors arising from diffusional encounters
%   with immobilized filaments}.
% \newblock {\em Phys. Rev. E}, 69(6):061911, June 2004.

% \bibitem{AbramovitzI.1972}
% M~Abramowitz and I~Stegun.
% \newblock {\em {Handbook of Mathematical Functions: with Formulas, Graphs, and
%   Mathematical Tables}}.
% \newblock Dover Publications, 1965.

% \bibitem{Arkhincheev1991}
% VE~Arkhincheev and EM~Baskin.
% \newblock {Anomalous diffusion and drift in a comb model of percolation
%   clusters}.
% \newblock {\em Zh. Exper. Teor. Fiziki}, 300(December 1990):161--165, 1991.

% \bibitem{Arkhincheev2002}
% V~E Arkhincheev.
% \newblock {Diffusion on random comb structure: effective medium approximation}.
% \newblock {\em Physica A}, 307(1–2):131--141, June 2002.

% \end{thebibliography}

\end{document}